# Application of Machine Learning to Performance Assessment for a class of PID-based Control Systems

Patryk Grelewicz, Thanh Tung Khuat, *Member, IEEE*, Jacek Czeczot, Pawel Nowak, Tomasz Klopot and Bogdan Gabrys, *Senior Member, IEEE*

***Abstract*— In this paper, a novel machine learning derived control performance assessment (CPA) classification system is proposed. It is dedicated for a wide class of PID-based control industrial loops with processes exhibiting dynamical properties close to second order plus delay time (SOPDT). The proposed concept is very general and easy to configure to distinguish between acceptable and poor closed loop performance. This approach allows for determining the best (but also robust and practically achievable) closed loop performance based on very popular and intuitive closed loop quality factors. Training set can be automatically derived off-line using a number of different, diverse control performance indices (CPIs) used as discriminative features of the assessed control system. The proposed extended set of CPIs is discussed with comprehensive performance assessment of different machine learning based classification methods and practical application of the suggested solution. As a result, a general-purpose CPA system is derived that can be immediately applied in practice without any preliminary or additional learning stage during normal closed loop operation. It is verified by practical application to assess the control system for a laboratory heat exchange and distribution setup.**

## I. Introduction

IN modern industrial control systems, high control performance of low-level controllers is crucial for efficient process operation [1]. This high performance is usually ensured by proper design [2]-[3] and tuning [4] of the controllers, e.g. using virtual commissioning approaches [5]-[6]. However, it is reported by practitioners that the quality of the control usually degrades over time due to fluctuations of process dynamics (e.g. resulting from slow fouling), slow decrease in accuracy of sensors and actuators or periodical modifications in production operating conditions [7]. The latter can result from unpredictable changes in a source of raw materials, periodical variations of major process disturbances, etc. This category also includes cases when controllers that operate the process were not properly tuned at the stage of commissioning and resulting production losses are not visible and evident. These facts are confirmed in the literature where the performance of over 60% of control loops has been observed to be poor [8] and in the vast majority of cases such a poor performance has resulted from a bad tuning of the controllers [9]. Thus, periodical control performance assessment (CPA) becomes more and more important. It can be considered as inessential part of fault detection systems that play a very important role in modern industry [10] and whose application is necessary to meet the requirements of Industry 4.0 in terms of preserving the best process efficiency [11]-[12]. Poor control performance must be detected and appropriate actions (e.g. appropriate controller retuning) must be taken, which is not easy when hundreds or even thousands of closed loops simultaneously operate on the process.

Comparing the actual performance of a control system with its reference performance is the fundamental principle underpinning various CPA algorithms. For a wide range of applications, the proposed procedure should therefore give explicit assessment if the control performance is satisfactory or poor by assessing how close it is to the desired reference performance.

Many CPA algorithms have been developed over last decades based on more or less complex mathematical and statistical approaches and they have gained popularity in both academia [13]-[14] and industry [15]-[16]. Apart from general approaches, some dedicated solutions were also reported. In [17] authors derive CPA method that is an important part of Iterative Learning Control (ILC) algorithm for control of batch processes. Dedicated CPA methods can be also applied for the design of fault tolerant control. An example of such application for the fault-tolerant control of singular systems was reported in [18].

This work was financed in part by the grant from SUT - subsidy for maintaining and developing the research potential in 2021 (J. Czeczot, P. Nowak, T. Klopot), and by BKM grant (BKM-723/RAU3/2020) (P. Grelewicz) and co-financed by the European Union through the European Social Fund (grant POWR.03.02.00-00-I029) (P. Grelewicz). Calculations were done with the use of GeCONiI infrastructure (POIG 02.03.01-24-099).
*(Corresponding author: J. Czeczot)*
P. Grelewicz, J. Czeczot, P. Nowak and T. Klopot are with Silesian University of Technology, Faculty of Automatic Control, Electronics and Computer Science, Department of Automatic Control and Robotics, Gliwice 44100, Poland (e-mail: jacek.czeczot@polsl.pl) .
T.T. Khuat and B. Gabrys are with University of Technology Sydney, Faculty of Engineering and IT, Advanced Analytics Institute, New South Wales 2007, Australia.

The first group of CPA methods is based on performing a comparison between the current control performance and the best observed so far in terms of the variance of manipulating and process variables [19]-[22]. These methods are based on normalized indices and their interpretation is clear. However, there is no explicit classification if the control performance is acceptable or not and how much this performance can be improved. Additionally, results depend strongly on stochastic characteristics of the process disturbances that in practice are often unknown and time-varying. Thus, these CPA algorithms can be used for monitoring a degradation in the control performance but not for its absolute assessment. They require a long "learning time" to get reliable information about the "so far the best performance" which is not readily and easily available. Thus, they require an initial stage of collecting process data and then, they can detect degradation compared to the "so far the best performance" but they fail when this detected "best performance" is far from "the best achievable performance".

The second group of methods is based on deriving and using general control performance indices (CPIs) that can be calculated for certain deterministic properties of a control system like a set point tracking and/or disturbance rejection. Based on time responses, different CPIs can be proposed, such as settling time, maximum overshoot, absolute square error, etc. [7] and it has already been shown that there exists a correlation between their values and the variance-based performance measures [23]. An application of these CPIs has been suggested for quantitative comparison between different controllers and/or different tunings and a list of different CPIs is very long. However, they focus on very limited properties of closed loop response and there is a lack of general rules regarding the way of using them for an explicit CPA. Additionally, there are no general "reference values" of CPIs and these "reference values" are case-dependent and must be adjusted accordingly for each new case. Thus, when CPIs are used for CPA problems, this approach has the same limitations as the algorithms described in the above paragraph in terms of the initial "learning time" and detecting the difference between "so far the best performance" and "the best achievable performance".

The motivation for this research was to derive a general-purpose CPA and, in this paper, it is tackled by proposing a machine learning derived CPA classification system. So far, in the vast majority of cases, ML methods are used for developing performance assessment systems but only for explicit technological process, e.g. for smelting process of electro-fused magnesium furnace [24]. More general approach can be found in [25], where the application of the kNN method to evaluate the performance of PID control system is demonstrated. Multi-class SVM has been proposed in [26], where based on time response data, the ACF coefficient and statistical features are calculated indicating potential problems with control system.

Proposed CPA is much more general even if its application is limited to conventional PID-based control loops working on a broad class of processes exhibiting dynamical properties close to second order+delay time (SOPDT). In industrial practice, this limitation is not very strong because PID controllers are still the most frequently used in low-level control loops and the vast majority of industrial processes can be accurately approximated by SOPDT dynamics. The proposed CPA system is based on the predefined reference disturbance rejection response of control system subject to SOPDT parameters and reference PID tuning. The acceptable deviation of this response is defined and a training dataset is generated by systematically simulating and recording acceptable and not acceptable disturbance rejection responses together with a set of related CPIs calculated from these responses. Once generated, this training dataset is used to train machine learning (ML) based classifiers to find accurate mapping between the CPIs and the class label (i.e. if the quality of control is acceptable or not). As part of the analysis of the feasibility and accuracy of such a mapping and its usefulness in control settings, a comprehensive comparative analysis of a wide range of ML based classification algorithms and an assessment of useful discriminative information contained in the proposed set of CPIs have also been performed. A simulation based validation shows applicability of the proposed CPA procedure to the PID-based closed loop systems with processes exhibiting different dynamical properties. Finally, practical cloud-based implementation of this system for PLC-based control loop is presented and experimental results show practical applicability of the proposed concept and its implementation.

The major novelty of this paper results from introducing the general concept of a machine learning-based CPA system for a wide class of industrial control loops, easy to configure off-line to distinguish between acceptable and poor closed loop performance by determining the best (but also robust and practically achievable) closed loop performance based on very popular and intuitive closed loop quality factors. As a result, this system can be immediately applied in practice without any preliminary or additional learning stage during normal closed loop operation.

The rest of this paper is organized as follows. Section II presents the statement of the problem. The design of the CPA system is discussed in Section III with a detailed analysis of an ML approach for the classification of a control performance presented in Section IV. Both simulation studies and a practical verification are summarized in Section V. Finally, Section VI concludes the paper. The main body of the paper is also complemented with the supplemental materials that present more implementation and validation results details.

For better clarity, section VIII of supplemental material includes the list of used abbreviations (Table S.IX) and symbols (Table S.X).

## II. Statement of the Problem

This study concentrates on the design of possibly the most general CPA system dedicated to classifying the control performance of closed loop systems with a conventional PID controller shown in Fig. 1. The control goal is defined to keep the process output *y* at a set point *sp* by minimizing the control error $e = sp - y$ with an efficient rejection of external disturbances.

The concept of a CPA system is also shown in Fig. 1. It is based on a direct assessment of the load disturbance rejection occurring as a result of a closed loop system excitation with a step change of an artificially introduced load disturbance $\Delta d$. This procedure can be enabled manually on demand of a user or applied periodically by a supervisory control system on a predefined schedule. When the CPA procedure is enabled, the system monitors the process output to detect a steady state and then, the load disturbing step change $\Delta d$ is applied to the closed loop system and the resulting response of the process output is collected until this disturbance is fully rejected and a new steady state is detected. Then, the disturbing $\Delta d$ is canceled and the control system returns to its normal operation while the CPA system computes certain features of the collected response and classifies whether the control performance is acceptable (OK) or not acceptable (NOK).

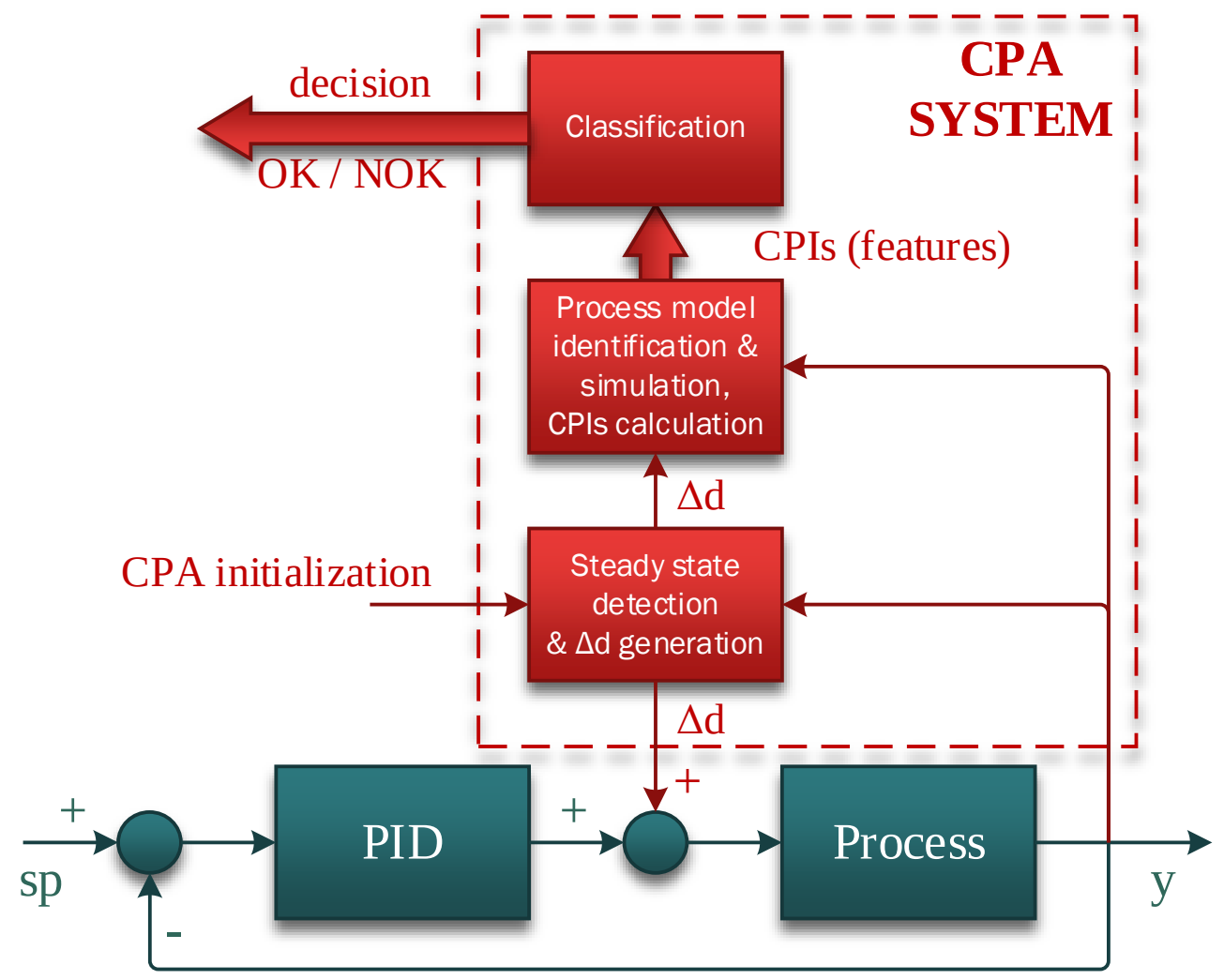

Fig. 1. PID-based closed loop system with schematic diagram of designed CPA system.

The proposed CPA system concentrates on assessing the disturbance rejection because in a process automation, vast majority of control systems are designed to provide effective disturbance rejection for a constant or rarely changed setpoint *sp*. Note that the concept of the proposed CPA procedure is similar to the self-tuning procedure widely applied for practical tuning of industrial PID controllers based on a built-in autotuning functionalities.

The assessment should be based on the purposely and carefully selected set of features of $\Delta d$ rejection response. These features should represent quantitative measures of the difference between predefined reference and current closed loop disturbance rejection responses. While a range of machine-learning methods can be applied to compare the predefined reference with the current closed loop response they require generating or collecting of appropriate, representative training data which is not a trivial task. Additionally, the CPA system should be effectively trained off-line so the assessment is possible without the necessity of any additional training for the target closed loop system. This procedure should not require any experience or expertise from the process operators, so the explicit assessment is essential.

It is also assumed that the suggested CPA system should be designed for an on-line assessment of the closed loop control systems consisting of a conventional PID controller that operates processes exhibiting possibly a wide range of dynamical properties.

## III. Design of CPA System

General concept of the suggested CPA system requires solving many practical difficulties.

### A. *Steady state detection and $\Delta d$ generation*

Practical steady state detection is an important issue and it is required in many practical situations, e.g. for an appropriate initialization of an autotuning procedure or for a signal-based process modelling. Many approaches have been proposed for this purpose and the most practically useful methods are: R-statistics-based method proposed in [27] and a simple but effective Increment Count Method (ICM) proposed in [28]. In this work, the latter method is used for a steady state detection.

The amplitude of $\Delta d$ should be adjusted to ensure a tradeoff between a sufficient process excitation and preventing from its inadmissible disturbing. In practice, this is a case-dependent value which must be selected based on the process dynamics and technological limitations.

### B. *Definition of reference disturbance rejection response*

The fundamental concept of the proposed CPA system for PID-based control systems comes down to the comparison between the so-called reference disturbance rejection response and the current one obtained after enabling CPA procedure. Thus, to ensure as high as possible generality of the CPA system, the reference disturbance rejection response must be predefined off-line and used for generating training datasets.

For a PID-based control system, the reference response depends on the PID tunings and parameters of the process dynamics. Thus, to ensure such high level of generality, it is required to assume the most general model of the process possible that ensures the trade-off between modeling accuracy and simplicity. Then, the reference PID tunings that ensure reference disturbance rejection response for a given process must be defined.

For modeling, it is assumed that the process can be precisely approximated by SOPDT dynamics with the following parameters: gain *k*, time constants $\tau_1 \geq \tau_2$ and delay time $\tau_0$. This assumption does not cause a very significant limitation as the majority of the industrial processes are self-regulating and stable. At the same time, contrary to very popular FOPDT (First Order+Delay Time) approximation, SOPDT model provides more precise approximation of higher order process dynamics. SOPDT model parameters can be easily computed from the process step response [28] but also from the closed loop rejection of intentionally applied load disturbance $\Delta d$ when current PID tunings and $\Delta d$ amplitude are known.

In practice, SOPDT time constants $\tau_1 \geq \tau_2$ and delay time $\tau_0$ can take positive but unlimited values and process gain can be also unlimited. Thus, appropriate scaling is suggested based on [29] and when this is performed the SOPDT approximation is described by normalized (unitary) gain and two relative dynamical parameters $L_1 = \tau_0/(\tau_1 + \tau_0)$ and $L_2 = \tau_2/\tau_1$. Both parameters $L_1$ and $L_2$ are limited between the values of 0 to 1 regardless of the values of the real SOPDT parameters. Additionally, the proposed CPA system is derived for SOPDT processes with additionally limited values of $L_1 \in [0.1, 0.6]$ and $L_2 \in [0.1, 1.0]$. These limitations include processes, for which application of PID controller is practically justified. For $L_1 > 0.6$, delay time is dominant and more advanced control strategies are suggested. At the same time, for $L_1$, $L_2 < 0.1$, a conventional PI controller can be easily tuned and applied.

For a given SOPDT process defined by unitary gain and $L_1$, $L_2$ parameters, the reference disturbance rejection response can be determined by adjusting the reference PID tunings: gain $k_r$, integral constant $T_i$ and derivative constant $T_d$. Note that the designed reference response should be not only achievable for a PID controller operating on a given process but also the corresponding reference PID tunings should preserve practical requirements defined for the control system, such as its robustness.

The so-called reference tuning is always relative and case-dependent and in this work, it is based on Integral Absolute Error (IAE) calculated for a disturbance rejection after exciting closed loop system with $\Delta d$. For a fixed SOPDT process parameters $L_1$, $L_2$ and constant $\Delta d$, IAE value depends only on the PID tunings and can be calculated by simulation as:

$$IAE(k_r, T_i, T_d) = \int_0^{t_{max}} |e(t)| dt, \qquad (1)$$

where $t_{max}$ denotes transient time after applying $\Delta d$. Then, based on Eq. (1), the following three-dimensional and constrained optimization problem can be defined:

$$\begin{aligned} & \underset{k_r, T_i, T_d \in \mathcal{R}^+}{minimize}\ IAE(k_r, T_i, T_d) \\ & \quad subject\ to \qquad A_m \geq 2.5 \\ & \qquad\qquad\qquad\quad\ \ \phi_m \geq 60^{\circ} \end{aligned} \qquad (2)$$

where $A_m$ and $\phi_m$ denote the gain and phase margins, respectively, and are defined to ensure desirable robustness of the closed loop and consequently to prevent too aggressive tuning. This approach is widely used for deriving tuning rules for various control algorithms, e.g. [30], and numerical solving of Eq. (2) allows for deriving IAE-based optimal tunings with desired robustness that in this work is considered as reference PID tunings $k_{r,ref}$, $T_{i,ref}$, $T_{d,ref}$.

Defining limiting gain and phase margins as $A_m \geq 2.5$ and $\phi_m \geq 60^{\circ}$ makes this tuning rather conservative but also acceptable from a practical viewpoint because it ensures relatively high closed loop robustness. Note that it is used only as an example in this work. One can apply different PID tuning methods for deriving the desired reference tunings and reference disturbance rejection responses, starting with popular experimental methods and ending with advanced optimization-based methods.

### *C. Control Performance Indices (CPIs) as a set of CPA features*

In this work, it is assumed that the proposed CPA is based only on the values of the selected CPIs computed from the rejection response to the applied disturbance step change $\Delta d$.

There are many well-known CPIs such as settling time, maximum overshoot or integral indices. In practice they are mainly used for two purposes. The first one is to design control systems or derive tuning rules. In this case, these indicators represent technological requirements or constraints. The second purpose of using CPIs is for a comparison of the performance of different control systems. In this work, the additional application of CPIs is proposed to assess whether a given load disturbance rejection trajectory is sufficiently similar to a reference trajectory. It is easy to see that using a single CPI is not sufficient. To illustrate this issue, four load disturbance rejection responses for differently tuned examples of PID controllers with the same SOPDT process (denoted as CS1, CS2, CS3, CS4) are presented in Fig. 2. CS1 is assumed as the reference trajectory characterizing the desired closed loop performance. CS1, CS2 and CS3 provide the same settling time. CS1 and CS4 provide the same maximum peak. However, all these disturbance responses have distinctly different shapes and characteristics. This is due to the fact that a single CPI is able to capture only very limited properties of the dynamic response. As a result, a single CPI cannot give correct CPA but the key features of load disturbance rejection responses can be captured by a number of different complementary CPIs.

The question arises how many indicators are needed to completely capture the key features of the response of the system and what should they be? As part of the investigation we have therefore decided to define and evaluate a wide range CPIs, many of which are novel and not previously used in the literature, in order to ensure that no important information will be omitted.

In order to systematize CPIs selection process, the load disturbance rejection response (see Fig. 2) is divided into three stages. A dynamical behavior at the first stage (starting from the moment of applying $\Delta d$ to the moment when the maximum peak appears) depends rather on the process dynamics, delay time and initial action of the PID controller. The behavior at the second stage characterizes the effectiveness of dumping the maximum peak. Finally, at the third stage it can be seen how the closed loop system is driven to a steady state. Thus, intuitively, the proposed and selected CPIs should capture the key features of each distinct stage of the load disturbance rejection and the key features of the whole response. It is worth noting that this is an informal classification that only facilitates the creation of the CPIs list and many other classifications can also be applied.

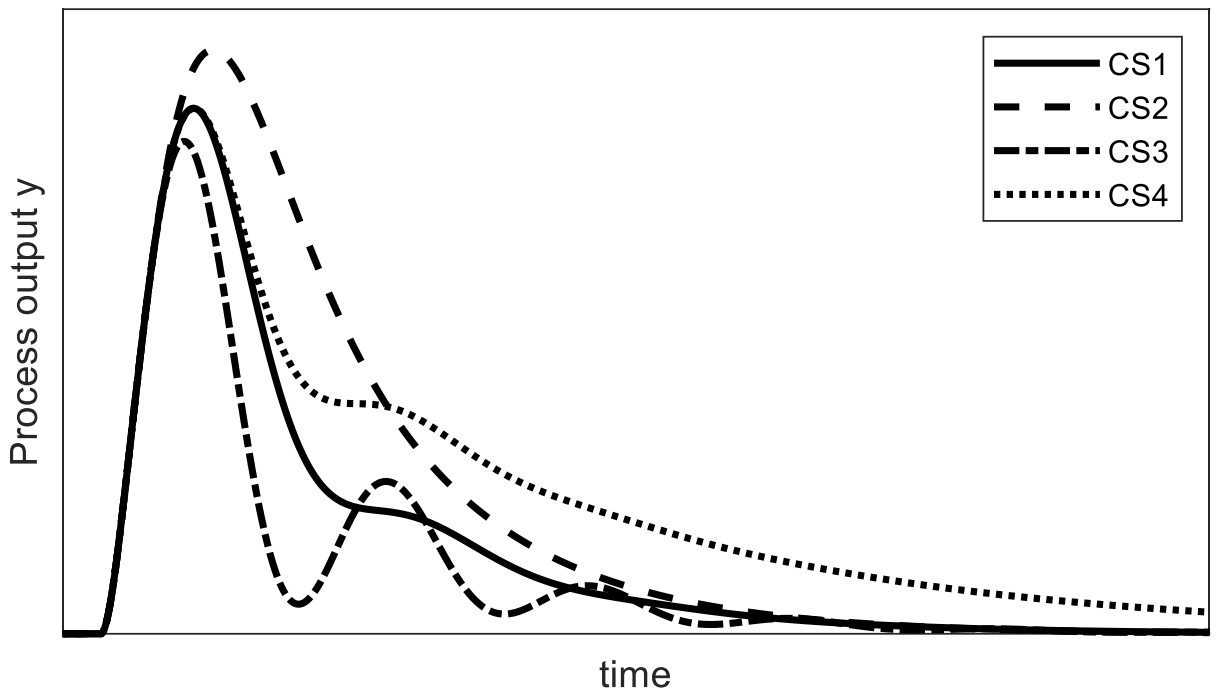

Fig. 2. Illustrative examples of responses of four differently tuned control systems to a step change of the load disturbance.

Following the above logic as a guidance in this work 30 different CPIs have been identified/proposed and ultimately selected for further evaluation. Their complete list is shown in Table S.I in supplemental material jointly with the graphical clarification of the meaning of some of them in Fig. S1. Twelve of the considered CPIs (highlighted with grey color) are very popular and commonly used by practitioners, i.e. maximum peak (F1), undershoot (F3), their ratio (F5), settling time (F7), various integral indices (F8 – F11), decay ratio (F15, F16) and finally, minimal and maximal values of response derivative (F28, F29).

CPIs F1, F28 and F29 describe the impact of initial controller action on the closed loop rejection response (the first overshoot after applying step change of $\Delta d$ and rate of output signal change). CPIs F3, F5, F15 and F16 focus on how the output signal varies after reaching the maximum peak and it attempts to quantify the aggressiveness level of the control action. CPIs F7 – F11 assess the overall closed loop transient time in correlation with the behavior of the control error *e*.

The other 18 CPIs are novel and their introduction is intended to capture much more nuanced dynamic characteristics of the assessed load disturbance rejection responses. Hence, these indices were mostly selected to complement the first 12 CPIs. Thus, the popular CPIs F1, F3 and F5 were respectively extended by F2, F4 and F6 to capture time domain features. They indicate the moments when overshoot and undershoot appear and their ratio is also captured. The absolute integral error index F8 was extended, resulting in F12 – F14, which are calculated according to different parts of dynamical response described by the sign of the control error *e*. These CPIs, jointly with other suggested integral indices F18 – F20, give more accurate information about overall properties of the response and its parts for positive and negative values of control error *e*. Settling time F7 was extended into F17, F24 – F27, where more key moments of response in time domain are detected.

The exceptions are indicators F21-F23, which were introduced to fully describe the first peak of the time response. They give information about the initial controller action (F21) and how effectively the first peak is dumped (F22). The following sections show that these features have serious impact on the performance of the whole CPA system.

All considered CPIs, therefore, define features of the assessed control system and they are computed from the applied disturbance rejection step response. The proposed list of CPIs was analyzed and some preliminary conclusions can be drawn:

- The proposed CPIs do not require high computational and memory resources for calculations. However, the derivative-based indices can be problematic to calculate in the presence of measurement noise for real process data. In this case, some additional filtering should be provided.
- Some indicators do not provide a straightforward assessment. For example, a long settling time can indicate too conservatively tuned control system with sluggish response or on the contrary, too aggressive tuning with oscillatory character (see Fig. 2).
- Some of these CPIs are not independent, e.g. control system with a long maximum peak time (F2) will probably also have a long rise time (F21). In addition, many CPIs are computed as ratios of other CPIs, so one can expect the correlation between them (e.g. F5, F14, F15, F23). However, it is worth emphasizing that these ratio-based CPIs are invariant for process parameters, which is promising in terms of their potential robustness without a need for scaling of the closed loop response subject to process dynamics.

Based on this analysis, preliminary intuitive selection of CPIs could be made for their suitability for the defined CPA problem. However, at this stage it was decided to use all of them. The possibility of potential reducing the number of indicators in order to avoid redundant information will be presented later in this paper.

## IV. Machine Learning Approach to Classification Models

The CPA problem defined in section II is proposed to be tackled and solved by designing a binary classifier based on a supervised machine learning (ML) approach. The use of a binary classifier ensures the explicit assessment of the control performance, i.e. if the control performance is satisfactory (OK) then the dynamic response is expected to be similar to its reference or poor (NOK) where the dynamic behavior is different. This concept is based on the thesis that a sufficiently large number of different CPIs defined in section III.C and capturing diverse, but key features of the dynamical load disturbance rejection response can provide consistent and useful information for such classification.

### A. *Generation of training and validation datasets*

The basis for deriving any classifier using machine learning approaches is the accessibility to training and validation datasets. In this case, it is assumed that after off-line training, the designed classifier should be ready for immediate application to operating PID-based control systems for their CPA. Thus, the stage of on-line training based on continuous observations of the behavior of control system under consideration is intentionally omitted. Off-line training should result in a properly designed classifier that does not require any additional training based on new process data.

Each time when the CPA procedure is enabled, load disturbance rejection response of the closed loop system in the presence of the applied $\Delta d$ step change is collected and this data is used for SOPDT process modeling. Thus, even if process dynamics varies subject to different reasons, at a given moment the CPA is made for a PID controller with given tunings and for an instantaneous SOPDT approximation of a given process.

Such an approach requires careful generation of both training and validation datasets. The basis for this generation is the reference load disturbance rejection trajectories computed by optimization (2) for a large and representative set of different SOPDT processes defined by $L_1$, $L_2$ dynamical parameters in the assumed ranges. For this purpose, the assumed ranges of $L_1 \in [0.1, 0.6]$, $L_2 \in [0.1, 1.0]$ variability were covered by a mesh of equidistant points with $\Delta L_1 = \Delta L_2 = 0.1$ so the boundary and internal points of this mesh represent 60 evenly distributed SOPDT processes. For each of them, reference PID tunings were derived by solving optimization problem (2). Then, based on the spline interpolation between reference PID tunings determined for neighboring mesh points, interpolated reference PID tunings were calculated for any combination of $L_1$, $L_2$ within the assumed ranges. This approach is considered to be sufficiently accurate and it allows for an approximate derivation of the reference PID tunings for each considered SOPDT process. However, if a higher interpolation accuracy is required, this mesh can be denser and the procedure can be easily repeated.

The control performance of a given control system should be assessed as OK, when its disturbance rejection response is similar to the reference one. That is why, more different responses of this closed loop system that are close to the reference response should be generated, covering the acceptable region of satisfactory control performance. For this purpose, reference PID tunings of any considered control system can be modified and corresponding disturbance rejection response can be computed by simulation. The modification was made by multiplying each reference PID tuning parameter ($k_{r,ref}$, $T_{i,ref}$, $T_{d,ref}$) by a random numbers $a_1$, $a_2$ and $a_3$:

$$\begin{aligned} k_{r,lab} &= a_1 k_{r,ref} \\ T_{i,lab} &= a_2 T_{i,ref}\ , \\ T_{d,lab} &= a_3 T_{d,ref} \end{aligned} \tag{3}$$

with a normal distribution N(1, 0.0225). Depending on a degree of this modification, one can obtain a control system of acceptable (OK) or not acceptable (NOK) control performance that can be included in the training and validation datasets. For each response, all 30 suggested CPIs are computed and their values form a feature vector representing the description of the response of the considered control system (i.e. they form a sample for the ML algorithms).

Subject to control performance, the binary labelling of each sample as OK or NOK is based on two criteria:

1) ± 10% acceptable deviation from the gain and phase margin computed for the control system under consideration, comparing to $A_{m,ref}$, $\phi_{m,ref}$ values characterizing the benchmark control system for corresponding $L_1$, $L_2$,
2) predefined normalized distance $e_{dist}$ between disturbance rejection responses for the control system under consideration $e_{lab}$ and reference $e_{ref}$ for given $L_1$, $L_2$:

$$e_{dist} = \frac{\int |e_{ref} - e_{lab}| dt}{\int |e_{ref}| dt}. \tag{4}$$

The control system under consideration is labelled OK if its gain and phase margin fall within the assumed range and $e_{dist} < 0.1$. Otherwise, it is labelled as NOK. This $e_{dist}$ threshold was adjusted experimentally based on preliminary studies which ensures that almost 96% of the control systems that meet this threshold, also meet required gain and phase margins. However, this value can be increased if greater deviation from reference response is acceptable as OK.

The training dataset was generated by selecting 60 000 control systems (samples) determined for random values of pairs $L_1$, $L_2$ within their assumed ranges and randomly modified reference PID tunings (3). It was ensured that for this training dataset, a half of the samples had to be selected from those labelled OK and the other half from the NOK class.

An example of the training dataset with the separation between OK and NOK ranges is graphically presented in Fig. 3 where green dots represent OK cases and red dots are NOK. For clarity, $A_{m,norm}$ and $\phi_{m,norm}$ respectively denote normalized distances of gain and phase margins and thus, their acceptable deviations are transformed into [-1, 1] range.

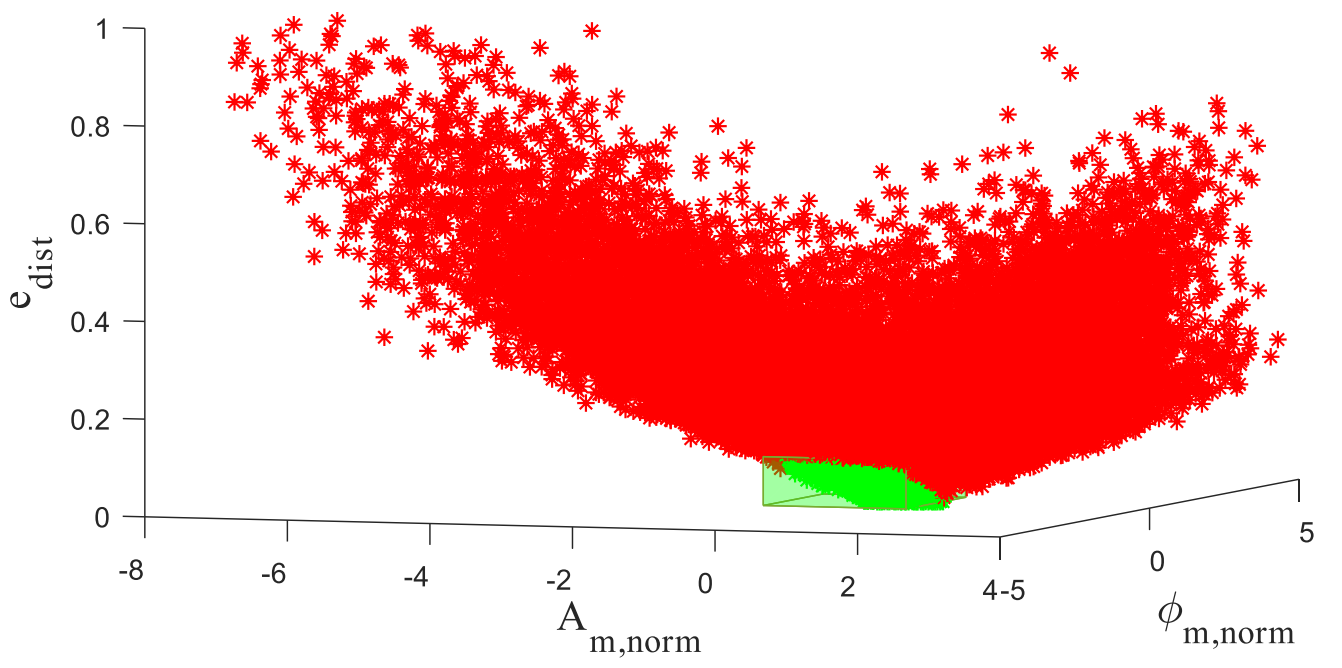

Fig. 3. Graphical representation of exemplary training dataset. OK and NOK performance is marked with green and red colors, respectively. Green box represents assumed range of OK performance.

The validation dataset was generated in the same way as training dataset (though completely independently for other random combinations of values of $L_1$ and $L_2$ within their ranges) but only 10 000 samples (control systems) for this dataset were selected. It was also ensured that a half was selected from those labelled OK and the other half from NOK. A feature vector for each sample was computed in the same way as for the training dataset and its labelling was also based on the same procedure.

### *B. Performance assessment of classification models*

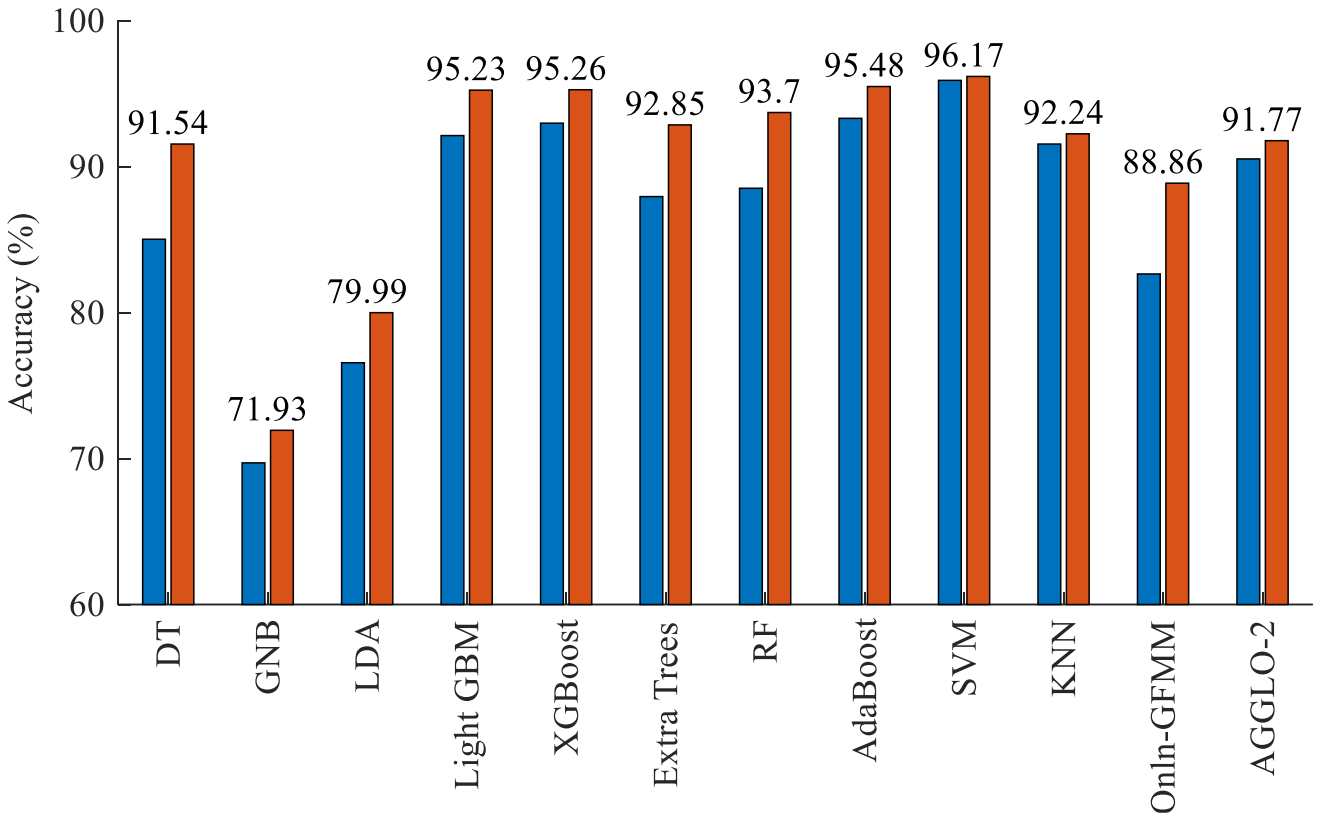

Fig. 4. Classification accuracy for considered classifiers obtained for validation dataset. Comparison between using popular 12 CPIs (features) and all 30 considered CPIs (features), both for training and validation.

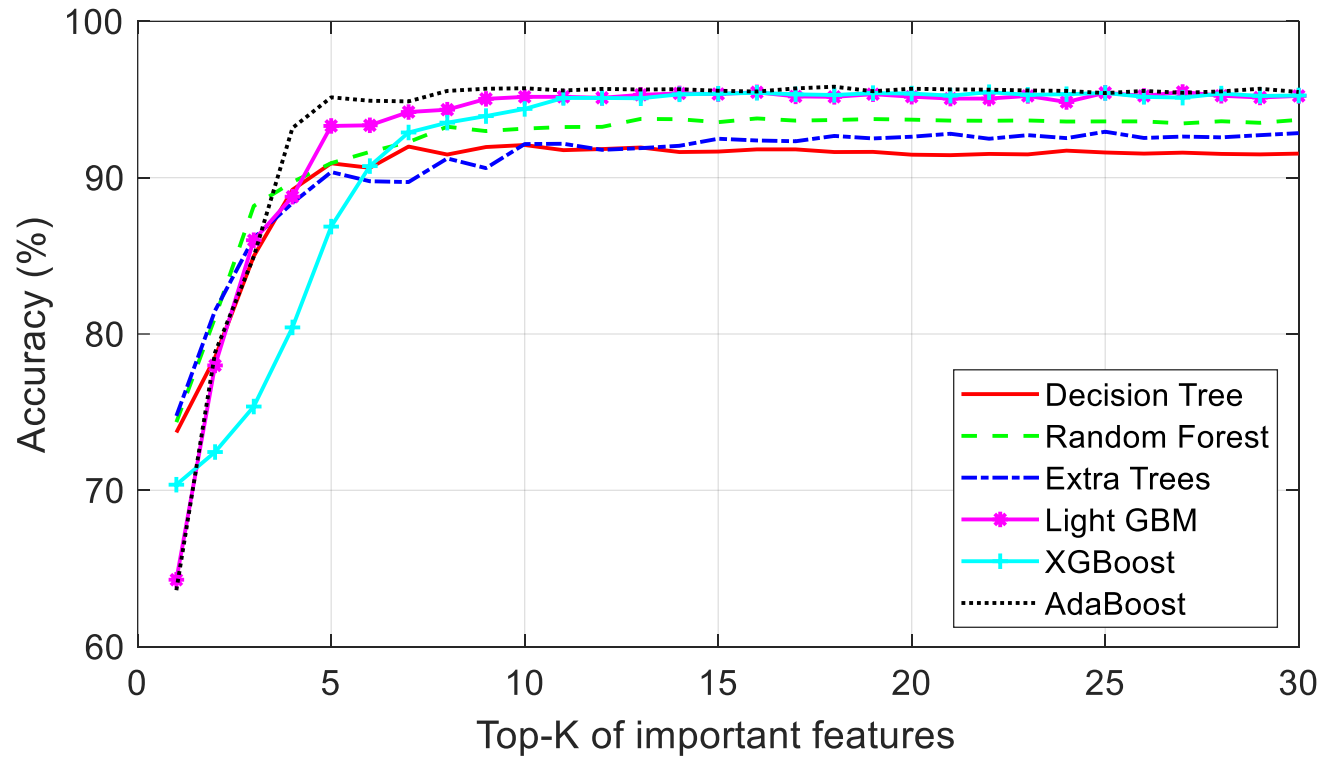

Fig. 5. Accuracy of tree-based learning models on the validation dataset using only top-k of the most important features.

Based on the training and validation datasets with 30 CPI features derived as described above, the classification performance of various machine learning algorithms for the considered CPA problem was assessed. Different types of classifiers were selected, ranging from the simple to complex but interpretable models such as Gaussian Naïve Bayes (GNB) [31], Linear Discriminant Analysis (LDA) [32], K-nearest Neighbors (KNN) [33], Decision Tree (DT) [34] and General Fuzzy Min-Max Neural Network trained by an online learning algorithm (Onln-GFMM) [35] or an agglomerative learning algorithm (AGGLO-2) [36], to less transparent but powerful classifiers including kernel-based methods such as Support Vector Machines (SVM) [37] and tree-based ensembles such as Light Gradient Boosted Machine (Light GBM) [38], Extreme Gradient Boosting (XGBoost) [39], Adaptive Boosting (AdaBoost) [40], Extremely Randomized Trees (Extra Trees) [41], and Random Forest (RF) [42]. Apart from GNB and LDA, hyper-parameters of the other models were tuned using random search with the maximum of 100 iterations and 5-fold cross-validation to find the best settings in given ranges as shown in Table S.II in the supplemental material.

Fig. 4 shows the classification accuracy for these classifiers on the validation dataset. Note that nine models achieved over 91% accuracy, and the best model, i.e., SVM, can achieve more than 96% accuracy. This figure additionally shows a comparison with the case when training and validation is based only on 12 most popular CPIs features. Note that in vast majority of the cases, the classification accuracy drops significantly, which clearly justifies extending the CPIs list to the 30 suggested features. As will also be illustrated and discussed later, a suitable combination of a subset of newly introduced and some of the well-known CPIs provides the best and most robust discriminative performance for different classifiers.

It can be seen that simple linear classifiers like GNB or LDA cannot reach 80% accuracy on the considered validation dataset. The best performances were observed for other non-linear models. These results indicate the decision boundary between samples of OK and NOK classes are of significantly non-linear nature and cannot be effectively captured by linear decision boundaries of GNB or LDA. As a result, non-linear classifiers were found to be the most appropriate for the CPA classification problem. It can be also noted that the use of complex but interpretable models such as DT, AGGLO-2, or KNN can result in quite good and competitive classification results compared to the other black-box complex models such as SVM or tree-based ensemble models. However, the best performance was usually achieved by using powerful non-linear classifier such as SVM or non-linear kernel and boosted ensemble classifiers, i.e., Light GBM, AdaBoost, and XGBoost.

Although the classification accuracy of fuzzy-based models such as Onln-GFMM and AGGLO-2 was lower than SVM or tree-based ensemble models, a strong argument for the use of these models is that their membership functions can be used to assess how close or far away from the acceptable and non-acceptable control performance boundary each of the classified samples is. This information can be useful to assess the effectiveness of CPA algorithms for monitoring the degradation of controllers in a dynamically changing environment and decide right times to retune the controllers. This opens an interesting research direction for future studies.

For the tree-based models, one of their interesting characteristics is the ability to extract individual CPIs importance scores. Given these importance scores for each tree-based model, the same classifiers were trained using only the top-k of the most important features, with k ranging from 1 to all 30 features. Feature ranking and classification performance of classifiers on subsets of the most important features are given in Table S.IV in the supplemental material. Fig. 5 summarizes the accuracy of these tree-based models on different subsets of the top-k of important features.

It can be observed that the accuracy of tree-based learners using from 8 to 15 of the most important features can achieve nearly equal or even better performance on the validation set compared to the case of using all 30 CPIs. This result poses a question of the optimal subset of CPIs which can be used in practice to attain the best classification performance of CPA systems instead of employing all of the proposed features.

While noting that substantially smaller set of features can be effectively used, as highlighted in Table S.IV in the supplemental material, the subsets may be different for different classifiers. Identifying a robust, minimal subset of discriminative features (i.e. CPIs) is out the scope of the current study and will be analyzed in more details in the future research.

Nevertheless, to provide further insights of what such reduced set of the CPIs may entail we will now analyze the top 10 CPIs with which the AdaBoost (the best performing algorithm in Fig. 5) algorithm obtained the best classification performance (see Table S.IV in supplemental material). These 10 CPIs are a mixture of more traditional indices and a number of the proposed in this study CPIs. As we can see, top two of them (F30 and F23) are the newly proposed ones and jointly with F1, F28 and F29, they mainly describe the properties of the first peak of the closed loop disturbance rejection response while F17 directly indicates the moment of time when this first peak appears.

Partially, F3 and F20 also relate to the first peak but they mainly inform about the properties of undershoot that may appear in some cases. F9 and F14 refer to the entire shape of the closed loop rejection response by quantifying integral (square or absolute) error and ratio of periods of time when control error has a negative and positive value.

Once again note that these properties are not sufficiently described by a single CPI. For example, rising and falling of the first peak are described by F23 and F30 but even if by intuition they seem to be highly correlated, they both have a strong impact on classification accuracy because the order in Table S.IV indicate their greatest importance. It is also worth noting that a large group of CPIs is calculated as a ratio between other CPI (F14, F20, F23, F30). Even if F30 is a ratio between F28 and F29 with the greatest importance, both F28 and F29 also play an important role in the construction of CPA classifier because they supplement the ratio-based F30.

Summarizing, it seems that the properties (shape, rising and falling times, etc.) of the first peak of the disturbance rejection response jointly with the description of the potential undershoot play the most important role in assessing the control performance.

In the next section, the effectiveness of learning models on simulation based and real process data is further assessed and discussed.

## V. Simulation Validation of CPA System

This section presents the results of the CPA performance based on SVM classifier selected due to its highest accuracy amongst all evaluated classifiers as reported in the previous section.

### *A. Validation for SOPDT processes*

Simulation based validation of the proposed CPA system was made for the selected SVM classifier but the classification accuracy obtained for the other classifiers for the simulation data is also shown in Table S.V in the supplemental material.

Simulation based validation was divided into two stages. At the first stage, initial validation was carried out by simulating the control systems with two different fixed SOPDT process respectively defined by ($L_1$ = 0.4, $L_2$ = 0.5) and ($L_1$ = 0.3, $L_2$ = 0.9). For each process, the testing dataset was generated by applying 35 different PID tunings based on FOPDT approximation of the process step response and arbitrarily selected from [43]. Thus, both testing datasets consist of 35 samples, each sample representing a different PID tuning method for the same SOPDT process.

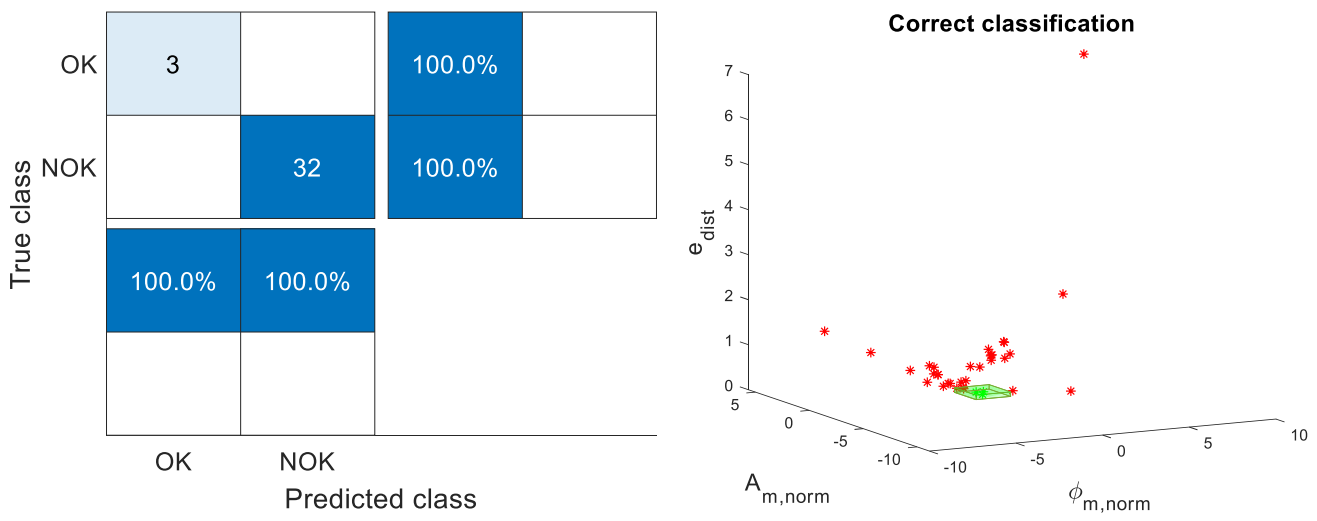

Fig. 6. (Left) Confusion matrix obtained for SVM classifier and test dataset. (Right). Graphical presentation of testing dataset, according to gain and phase margins and $e_{dist}$. SOPDT ($L_1$ = 0.4, $L_2$ = 0.5).

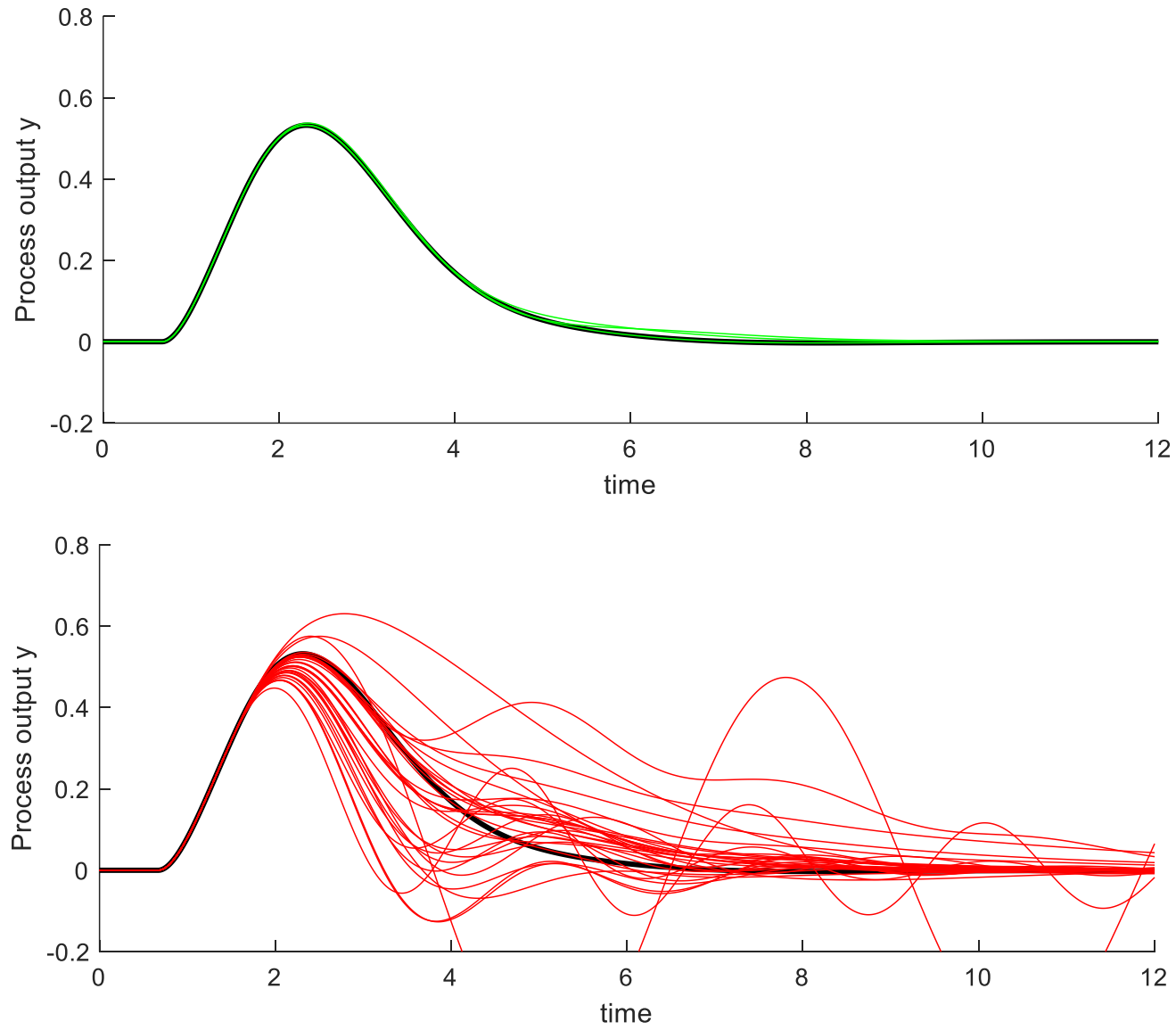

Fig. 7. Comparison of reference response (thick, black plot) with testing control systems classified as OK (green upper plots) and NOK (red lower plots). SOPDT ($L_1$ = 0.4, $L_2$ = 0.5).

Fig. 6 shows the classification accuracy for the testing dataset representing SOPDT with ($L_1$ = 0.4, $L_2$ = 0.5), which for this case is perfect (i.e. 100%). Fig. 7 shows the disturbance rejection responses for each sample from this testing dataset. Note that those classified as OK are very similar to the reference response of the control system with the considered SOPDT process. At the same time, responses classified as NOK are far from it and some of them are surely not acceptable in practice.

For the second testing dataset representing SOPDT process with ($L_1$ = 0.3, $L_2$ = 0.9), one set of PID tunings leads to unstable behavior. The classification accuracy shown in Fig. 8 is still very high but not perfect. One control system was misclassified as NOK while in accordance with the labelling methodology described in Section IV.A, it should be classified as OK. Fig. 9 shows its disturbance rejection response. However, graphical representation of this testing dataset shows that the misclassified sample is very close to the border of NOK region. It is obvious, that in practice, the accuracy of classifiers will not be perfect, especially when testing samples are relatively close to the border between OK and NOK classes. To further distinguish between the cases close to the decision boundaries and provide additional information beyond the class labels, the membership functions of GFMM classifiers can be used and will further be explored in the follow up studies.

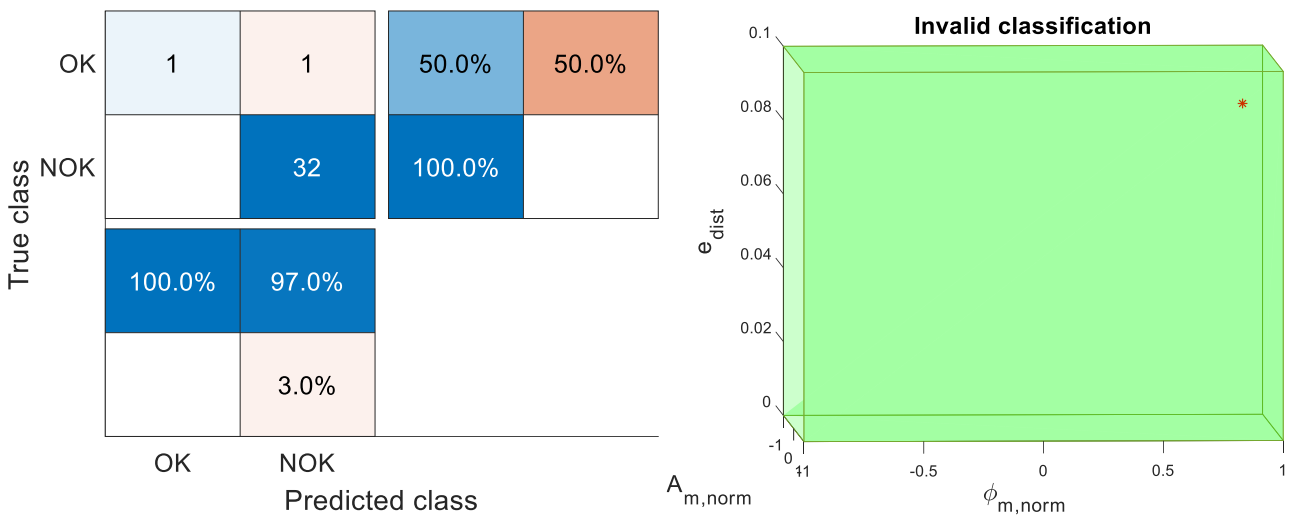

Fig. 8. (Left) Confusion matrix obtained for SVM classifier and test dataset. (Right). Graphical presentation of testing dataset, according to gain and phase margins and $e_{dist}$. SOPDT ($L_1$ = 0.3, $L_2$ = 0.9).

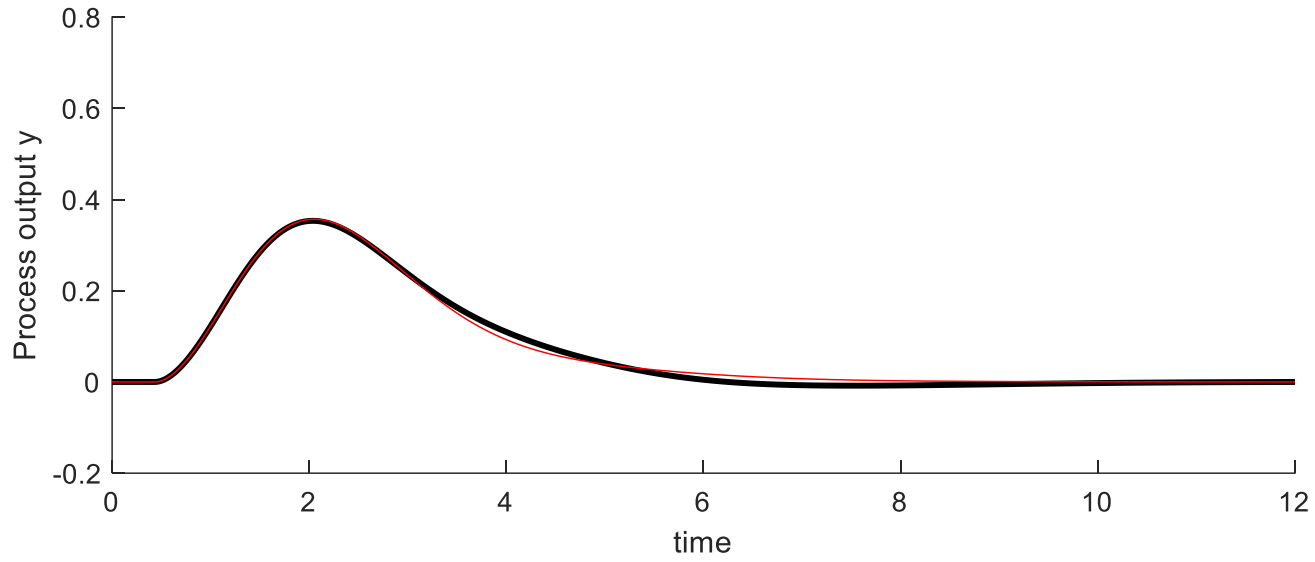

Fig. 9. Comparison of reference response (thick, black plot) with testing control system misclassified as NOK (red plot). SOPDT ($L_1$ = 0.3, $L_2$ = 0.9).

## *B. Comparison with existing methods*

The suggested CPA system was compared with other existing CPA methods. Based on disturbance rejection response data, the performance can be assessed with R Index [44], Idle Index [45], Area Index [46] and Load disturbance Rejection Performance (LDR) Index [47]. These indices are more general than individual CPIs and they can be applied for more precise assessment of control performance based on their values shown in Table S.VI in supplemental material.

Assessing procedure was similar to one applied for the testing of the suggested CPA system. Based on the generated simulation datasets (for $L_1$ = 0.4, $L_2$ = 0.5 and $L_1$ = 0.3, $L_2$ = 0.9), CPA indices selected for comparison were calculated and the results are presented in supplemental material in Tables S.VII and S.VIII. They are color-coded according to Table S.VI, where OK and NOK assessment is highlighted with green and red colors, respectively. For better clarity, the results are also presented in graphical form in Fig. S4. One can notice, that the application of CPA indices selected for this comparison do not ensure distinguishing between OK and NOK samples. Thus, it is not possible to correctly assess the control performance based on individual CPA indices. In [48] authors suggest application of both Idle Index and Area Index for more precise assessment, however even focusing on all of the selected indices, without any systematic framework, does not ensure proper assessment.

One can notice, that for ($L_1$ = 0.4, $L_2$ = 0.5), there are several process responses (no 16, 27, 28 and 29), which are assessed as OK by all of the CPA methods selected for comparison, but based on criteria chosen for deriving proposed CPA system, the performance is poor (NOK). These process responses are presented in Fig. S3 and their dynamic behavior is different from predefined reference. Additionally, one can notice even oscillatory behavior, which is not acceptable from practical viewpoint.

The suggested CPA system was also compared with Harris index [49], which is a more complex method for CPA. Harris Index requires stochastic-type disturbance and for this purpose, several steps of load disturbance with different amplitude were applied to the assessed control systems (for $L_1$ = 0.4, $L_2$ = 0.5 and $L_1$ = 0.3, $L_2$ = 0.9) with selected tunings. Note that in this case, much more aggressive excitation must be applied to the closed loop system, comparing to a single step change of load disturbance required for the suggested CPA system. The results of the assessment with Harris index are also presented in supplemental material in Tables S.VII and S.VIII.

Harris index is normalized from 0 (worse performance) to 1 (best performance) and it compares the performance of actual control system with the performance which can be achieved for the minimum variance controller. However, in practice, the minimum variance controller is not applicable, thus it is impossible to reach unitary value of Harris index. It is not clear what value of Harris index is achievable for PID-based closed loop system so in practice, its reference value is unknown. Thus, the explicit assessment based on Harris index can be a challenging task, due to its ambiguity.

## *C. Validation for higher order processes*

The second stage of simulation based validation was carried out for two processes whose dynamical properties are significantly different from SOPDT and their SOPDT approximation was used only for CPA. Their dynamical properties are given by transfer functions taken from [50] with an additional supplementation of $G_2(s)$ with scalable delay time term:

$$G_1(s) = \frac{1}{(1+s)^{\alpha}}, \tag{5a}$$

$$G_2(s) = \frac{1}{(1+s)(1+\alpha s)(1+\alpha^2 s)(1+\alpha^3 s)} e^{-\alpha s}. \tag{5b}$$

Both transfer functions (5) can be parameterized by adjusting the value of $\alpha$ and Table I shows the selected processes considered for the validation of the CPA system. Note that the precise selection of $\alpha$ allows to obtain processes whose SOPDT approximations quite evenly cover the assumed range of $L_1$, $L_2$

TABLE I
SELECTED PROCESSES USED FOR CPA VALIDATION

| PROCESS ACRONYM | TRANSFER FUNCTION | $L_1$ | $L_2$ | FIG. NO. IN SUPPL. MAT.* |
|---|---|---|---|---|
| *P1* | $G_1$, $\alpha = 3$ | 0.27 | 1.0 | S5 |
| *P2* | $G_1$, $\alpha = 4$ | 0.41 | 1.0 | S6 |
| *P3* | $G_2$, $\alpha = 0.25$ | 0.24 | 0.28 | S7 |
| *P4* | $G_2$, $\alpha = 0.3$ | 0.28 | 0.33 | S8 |
| *P5* | $G_2$, $\alpha = 0.4$ | 0.37 | 0.5 | S9 |
| *P6* | $G_2$, $\alpha = 0.5$ | 0.49 | 1.0 | S10 |
| *P7* | $G_2$, $\alpha = 0.6$ | 0.53 | 1.0 | S11 |

(*) last column shown numbers of figures showing results for corresponding processes in section VI in supplemental material.

For each process, 20 different sets of PID tunings were selected representing 20 different control systems (samples). Some of them were based on well-known tuning methods [40] while the others were adjusted by the trial and error method to obtain the possibly highest control performance. Then, for each set of the PID tunings and each process, the same CPA system designed as described above was applied. It was operated with the applied load disturbance $\Delta d = 1$.

Detailed results of this stage of validation are presented in section VI in the supplemental material. For each considered process *P1 – P7* it can be seen that SOPDT model provides more precise approximation of dynamical properties in comparison with more popular FOPDT model. This observation additionally justifies the choice of SOPDT modelling as more precise and general. One can also note that for processes *P3 – P7* that are based on dynamics given by Eq. (5b), the reference disturbance rejection response of the real process is very close to the one obtained for the closed loop system with corresponding SOPDT approximation. For processes *P1 – P2* that are based on dynamics given by Eq. (5a), this similarity is lower but the shapes and major properties are still preserved.

When it comes to CPA results obtained for the suggested system, one can see that the accuracy of classification is very high. For each process, rejection responses classified as OK are close to corresponding reference rejection trajectory while those classified as NOK significantly deviate from it. Once again, due to accuracy of SOPDT approximation, for processes *P3 – P7*, responses classified as OK are very close to their reference. For processes *P1 – P2*, responses classified as OK are more different than their reference but still these differences are acceptable compared to cases of rejection responses classified as NOK. At the same time, even if these differences are more noticeable in comparison with processes *P3 – P7*, responses classified as OK form their own similar shape and in this sense, they form their own reference slightly different from those obtained for SOPDT approximations but still acceptable from the practical viewpoint.

## VI. EXPERIMENTAL VALIDATION

To further evaluate and strengthen the argument in support of the proposed approach, an experimental validation was performed based on the part of laboratory heat exchange and distribution plant shown in Fig. 10. Experiments were carried out for the electric flow heater with adjustable heating power $P_h$ within the range 0 - 100% of maximal power 12 kW. The water flows through the heater with the flow rate $F$ and temperature is measured at the heater inlet ($T_{in}$) and outlet ($T_{out}$). The control goal is defined to ensure that $T_{out} = T_{SP}$ (temperature setpoint) by manipulating heating power (manipulating variable). This process exhibits higher (above second) order dynamics with significant delay time, so its dynamical properties are different from SOPDT used for the training of the CPA system.

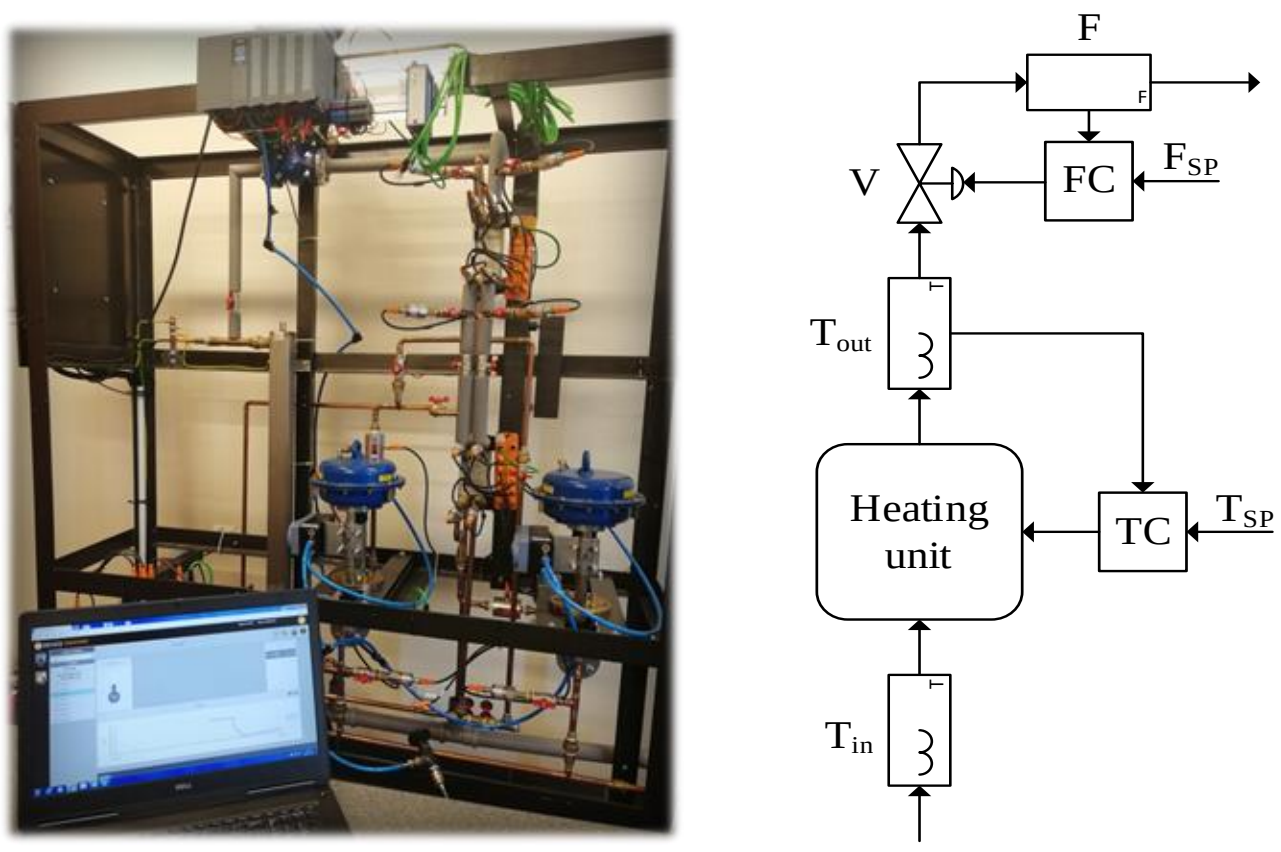

Fig. 10.The overview (left) and simplified diagram (right) of laboratory setup.

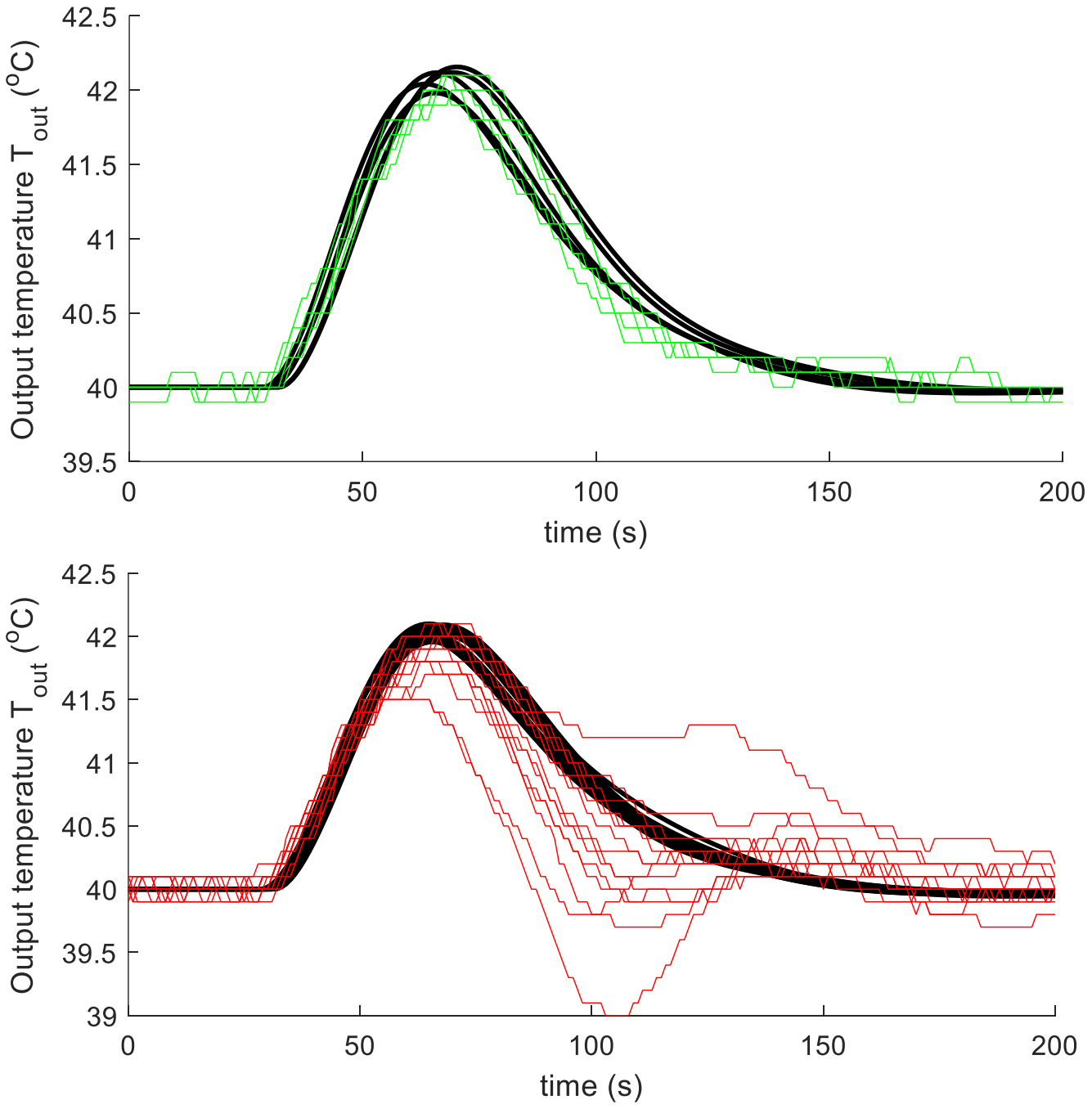

Fig. 11. Comparison of reference responses (thick, black plots) with testing control systems classified as OK (green upper plots) and NOK (red lower plots) obtained from laboratory setup.

Details of the practical cloud-based implementation of the proposed CPA system in the application to CPA of a PID controller implemented in Siemens S7-1500 PLC and operating the process are presented in section VII in the supplemental material.

For constant flow rate $F$ = 3.5 L/min, similarly to the second stage of simulation based validation, 20 different sets of PID tunings were selected to represent 20 different control systems (samples). Then, for each set of the PID tunings, a laboratory setup was operated, and the CPA procedure was executed. It was operated with the applied load disturbance $\Delta d = \Delta P_h = 10\%$.

The classification for 20 collected experimental rejection disturbance step responses are shown in Fig. 11. For the visualized measurement data, one can see a presence of the quantization resulting from limited sensor resolution. Note that in this case, corresponding reference responses are slightly different for each CPA experiment. It results from the fact that in practice it is impossible to obtain the same results even in the same conditions. Thus, for each CPA experiment, SOPDT approximation of the real disturbance rejection step response is slightly different.

The results show very high classification accuracy for the selected SVM model in the application to CPA of the real process exhibiting dynamics more complex than SOPDT. Rejection responses classified as OK are close to the corresponding reference rejection trajectories while those classified as NOK are significantly different and not acceptable in practice.

## VII. Conclusions

This paper introduces the concept of machine learning (ML) based CPA system and investigates its application to assess the performance of PID-based control loops operating processes that exhibit dynamics close to SOPDT. The proposed concept is based on fusion of up to 30 individual, diverse CPIs computed from the disturbance rejection step response of the assessed control system. These CPIs are used as input features to the ML based classification system. A comparative analysis of a wide range of different machine learning algorithms is presented and important conclusions are drawn in terms of potential reduction of a number of features required for an accurate classification.

Set of the considered CPI features consists of 12 very popular CPIs and 18 additional ones specifically proposed for this study. The classification accuracy and feature importance analysis showed that in general, these additional features provide more effective discriminative representation of properties of the assessed control systems. Thus, the results indicated that a relatively small subset of them can be used for an accurate assessment of the control performance if a load disturbance step change, required for their calculation, can be applied.

The proposed CPA system partially falls within the category of data-driven implementation of active fault detection systems [51]-[52] with model-based and signal-based approach. From that perspective, the considered degradation of control performance does not fall into the category of faults that require fast service (replacement) activities and its bad influence can be compensated only by periodical retuning of operating controller. However, the proposed CPA system can be included as a part of more complex fault detection, isolation and identification system that can suggest further actions (e.g. replacement of a sensor or actuator) when the controller retuning is no longer sufficient. Thus, the application of the proposed CPA system allows not only for improvement in the control performance by periodical controller retuning but it can also postpone the moment of replacing the partially worn out parts.

The proposed approach requires identification of SOPDT process parameters from the closed loop disturbance rejection response. Thus, it allows also for adding the functionality of retuning the PID controller. This possibility is beyond the scope of this work but readers should note that once SOPDT process approximation is known, it can be used for suggesting the PID controller tunings that provide the desirable control performance.

Promising results show that this concept can be extended to other classes of control systems, which are based on different (even advanced) controllers operating processes exhibiting different (even more complex) dynamics. At the same time, the proposed framework itself is general and flexible, which is shown by clock diagram presented in section III in supplemental material. After redefining some initial assumptions, this approach can be reconfigured to current needs and used for off-line designed of a new CPA system.

The proposed approach, with some indicated extensions forming our future research directions, can be also applied for the assessment of tracking properties of the operating control systems. The included example of practical implementation shows potential applicability and easy transferability of the proposed CPA system into the industrial practice.

# Supplemental Material for the Paper: Application of Machine Learning to Performance Assessment for a class of PID-based Control Systems

Patryk Grelewicz, Thanh Tung Khuat, *Member*, *IEEE*, Jacek Czeczot, Pawel Nowak, Tomasz Klopot and Bogdan Gabrys, *Senior Member*, *IEEE*

## I. Complete list of Used Control Performance Indices (CPIs)

The complete list of used CPIs with their short descriptions is presented in Table S.I. The most popular CPIs that are frequently used as control performance measures are highlighted with grey colour while the other CPIs are defined specifically for this work.

TABLE S.I
THE COMPLETE LIST OF CPIS

| Control Performance Index | Short description | Acronym |
|---|---|---|
| *MaxPeak* | Maximum value of dynamic system response | F1 |
| *MaxPeakTime* | The moment, when the maximum peak occurs | F2 |
| *MinPeak* | Minimum value of dynamic system response, absolute value | F3 |
| *MinPeakTime* | The moment, when the minimum peak occurs | F4 |
| *MinToMax* | The ratio of minimum and maximum peak | F5 |
| *MaxToMinTime* | The difference of time, when maximum and minimum peaks occur $MaxToMinTime = MinPeakTime - MaxPeakTime$ | F6 |
| *SettlingTime* | The moment, when the response of system is within the range of 1% of its steady state $\lvert e\rvert < 0.01$ | F7 |
| *IAE* | Integral Absolute Error $IAE = \int \lvert e\rvert dt$ | F8 |
| *ISE* | Integral Square Error $ISE = \int e^2 dt$ | F9 |
| *ITAE* | Integral Time Absolute Error $ITAE = \int t\lvert e\rvert dt$ | F10 |
| *IT2AE* | Integral Time Square Absolute Error $IT2AE = \int t^2 \lvert e\rvert dt$ | F11 |
| *IAEPos* | Integral Absolute Error calculated for positive values of system response $IAEPos = \int \lvert e\rvert dt, e > 0$ | F12 |
| *IAENeg* | Integral Absolute Error calculated for negative values of system response $IAENeg = \int \lvert e\rvert dt, e < 0$ | F13 |
| *IAENegToPos* | Ratio of *IAENeg* and *IAEPos* | F14 |
| *DecayRatio* | Ratio of maximum peak to second positive peak $DecayRatio = \frac{2^{nd}Peak}{MaxPeak}$ | F15 |
| *DecayRatioTime* | The difference between time, when maximum and second peaks appeared $DecayRatioTime = 2^{nd}PeakTime - MaxPeakTime$ | F16 |
| *PeakSettlingTime* | Difference between *SettlingTime* and *MaxPeakTime* | F17 |
| *TimePos* | The total amount of time, when the response of the system is positive $TimePos = \int dt, e > 0$ | F18 |
| *TimeNeg* | The total amount of time, when the response of the system is negative $TimeNeg = \int dt, e < 0$ | F19 |
| *TimeNegToPos* | The ratio of *TimeNeg* and *TimePos* | F20 |
| *RisingTime* | Rising time of the maximum peak, calculated as a time of reaching from 5% to 95% of *MaxPeak* | F21 |
| *FallingTime* | Falling time of the maximum peak, calculated as a time of reaching from 95% to 5% of *MaxPeak* | F22 |
| *RisingToFallingTime* | Ratio of *RisingTime* and *FallingTime* | F23 |
| *25%DistRejected* | The moment, when the response of system is within the range of 25% of *MaxPeak*, \|*e*\| < 25%**MaxPeak* | F24 |
| *50%DistRejected* | The moment, when the response of system is within the range of 50% of *MaxPeak*, \|*e*\| < 50%**MaxPeak* | F25 |
| *75%DistRejected* | The moment, when the response of system is within the range of 75% of *MaxPeak*, \|*e*\| < 75%**MaxPeak* | F26 |
| *ZeroCrossingTime* | The first moment, when the response of the system crosses the zero value | F27 |
| *MaxDiff* | Maximum value of the derivative of the dynamic response | F28 |
| *MinDiff* | Minimum value of the derivative of the dynamic response, absolute value | F29 |
| *DiffMaxToMin* | Ratio of *MaxDiff* and *MinDiff* | F30 |

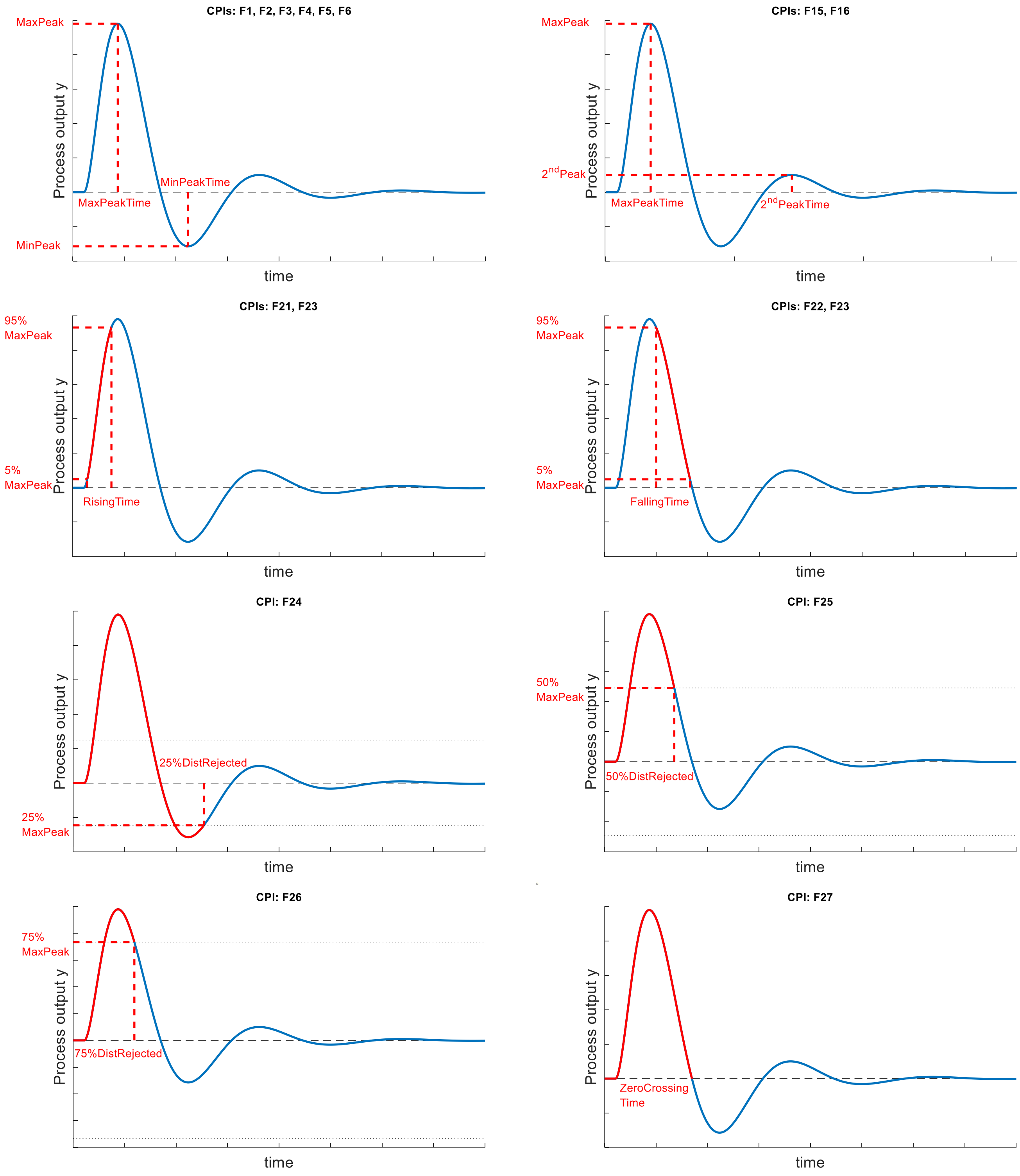

Fig. S1. Graphical interpretation of a set of chosen CPIs: *MaxPeak*, *MaxPeakTime*, *MinPeak*, *MinPeakTime*, *2$^{nd}$Peak* (for calculating *DecayRatio*), *2$^{nd}$PeakTime* (for calculating *DecayRatioTime*), *RisingTime*, *FallingTime*, *25%DistRejected*, *50%DistRejected*, *75%DistRejected*, *ZeroCrossingTime*.

## II. Hyperparameter optimization

The parameters of studied classification methods were obtained using a hyperparameter optimization approach described in the main manuscript. The results are presented in Table S.II, including the considered range and optimal value of each hyperparameter.

TABLE S.II
HYPERPARAMETER OPTIMIZATION RESULTS FOR STUDIED CLASSIFICATION ALGORITHMS

| Classification algorithm | Parameter | Range | Optimal value |
|---|---|---|---|
| Decision Trees | Max depth | [4, 20] | 19 |
| | Min samples per leaf | [4, 30] | 4 |
| Light GBM | Max depth | [4, 20] | 20 |
| | Min samples per leaf | [4, 30] | 12 |
| | Sampling rate | {0.3, 0.4, 0.5, 0.6, 0.7} | 0.4 |
| | % features used | {20%, 30%, 40%, 50%, 60%, 70%} | 70% |
| | Learning rate | {0.025, 0.05, 0.1, 0.2, 0.3} | 0.3 |
| | No of estimators | {30, 50, 70, 100, 150, 200} | 200 |
| XGBoost | Max depth | [4, 20] | 8 |
| | Sampling rate | {0.3, 0.4, 0.5, 0.6, 0.7} | 0.7 |
| | % features used | {20%, 30%, 40%, 50%, 60%, 70%} | 70% |
| | Learning rate | {0.025, 0.05, 0.1, 0.2, 0.3} | 0.2 |
| | Gamma | {0, 0.1, 0.2, 0.3, 0.4, 1, 1.5, 2} | 1 |
| | No of estimators | {30, 50, 70, 100, 150, 200} | 200 |
| Extra Trees | Max depth | [4, 20] | 20 |
| | Min samples per leaf | [4, 30] | 6 |
| | % features used | {20%, 30%, 40%, 50%, 60%, 70%} | 40% |
| | Sampling rate | {0.3, 0.4, 0.5, 0.6, 0.7} | 0.7 |
| | No of estimators | {30, 50, 70, 100, 150, 200} | 50 |
| Random Forest | Max depth | [4, 20] | 20 |
| | Min samples per leaf | [4, 30] | 6 |
| | % features used | {20%, 30%, 40%, 50%, 60%, 70%} | 40% |
| | Sampling rate | {0.3, 0.4, 0.5, 0.6, 0.7} | 0.7 |
| | No of estimators | {30, 50, 70, 100, 150, 200} | 50 |
| AdaBoost | Max depth | [4, 20] | 11 |
| | Min samples per leaf | [4, 30] | 12 |
| | No of estimators | {30, 50, 70, 100, 150, 200} | 150 |
| | Learning rate | {0.001, 0.01, 0.1, 0.2, 0.5, 1} | 0.1 |
| Support Vector Machines | Kernel | {'rbf', 'sigmoid', 'linear'} | rbf |
| | Gamma | {2^-15, 2^-13, …, 2^3} | 8 |
| | C | {2^-5, 2^-3, …, 2^15} | 512 |
| K-nearest Neighbour | K | {1, 3, …, 29} | 5 |
| Onln-GFMM | Maximum hyperbox size θ | {0.1, 0.15, …, 0.55, 0.6} | 0.1 |
| AGGLO-2 | Maximum hyperbox size θ | {0.1, 0.15, …, 0.55, 0.6} | 0.4 |

## III. Block diagram of suggested CPA system

Block diagram representing the stages of deriving the proposed CPA system is presented in Fig. S2. Note the generality of the suggested procedure resulting from its configurability at different stages. In this paper, example (and practically justified) configuring parameters are proposed nut there is a possibility of using different process models, different criteria of assessment and different set of features for the same procedure.

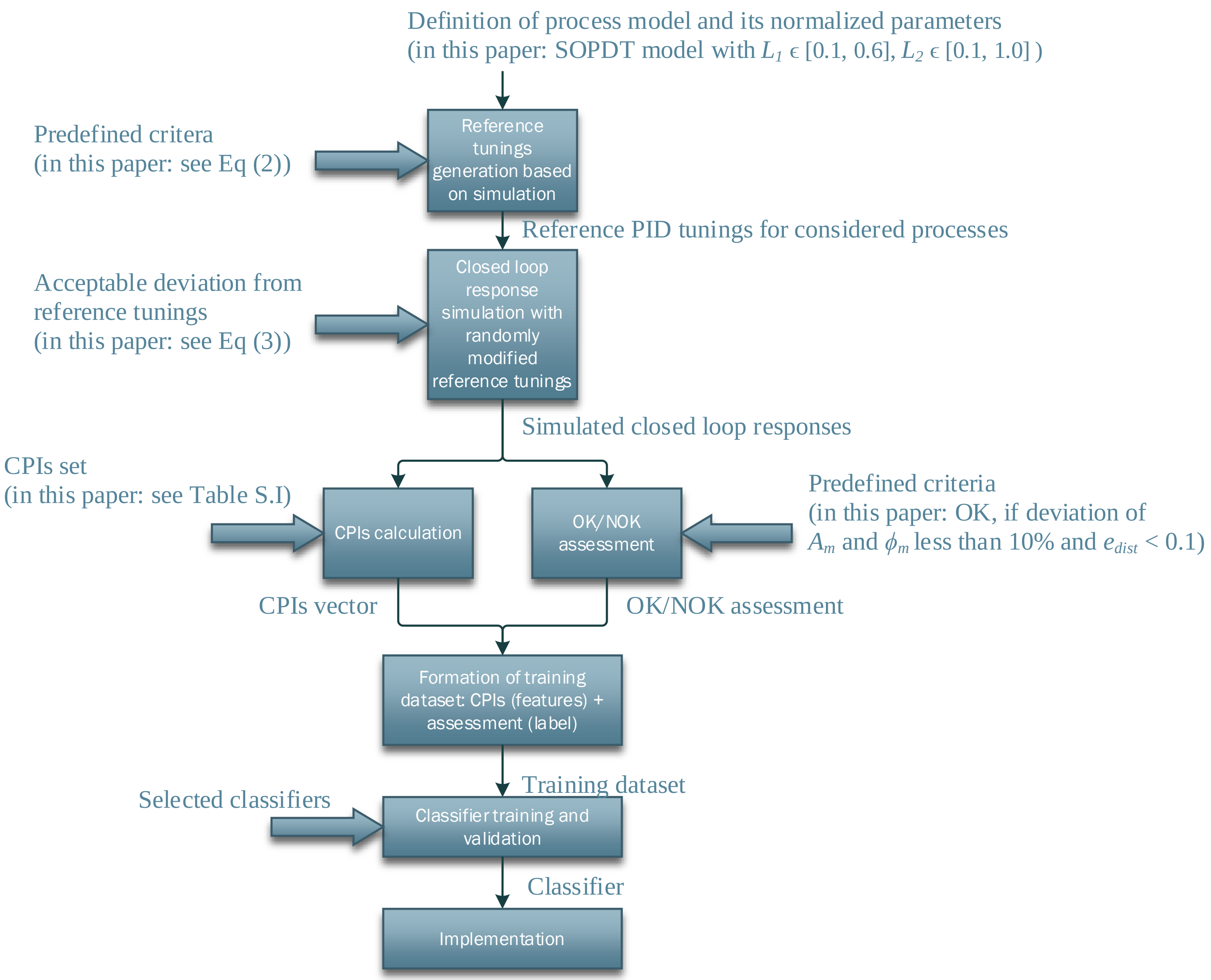

Fig. S2. Block diagram of general approach to deriving suggested CPA system.

## IV. Possibility of Feature Reduction

To check the possibility of feature reduction, correlation coefficients (Table S.III) and feature importance for tree-based models (Table S.IV) were calculated. The highly correlated groups of indices were colour-coded in Table S.III and Table S.IV. One can notice that the most important features in the vast majority of cases are the representatives of obtained colour-coded groups. What is more, the classification accuracy does not increase, when the number of features is higher than approximately 10 (Fig. 5 in the main paper). These results suggest that the number of effective CPIs can be reduced without any significant drop in classification accuracy. This issue will be studied in the future, as with a small number of relatively easily computable features, the overall computational complexity decreases and a type of the CPA system proposed in this work can be implemented directly in PLC, as a ready-to-use general-purpose function block.

TABLE S.III
Correlation coefficients calculated for each pair of CPI

| | F1 | F2 | F3 | F4 | F5 | F6 | F7 | F8 | F9 | F10 | F11 | F12 | F13 | F14 | F15 | F16 | F17 | F18 | F19 | F20 | F21 | F22 | F23 | F24 | F25 | F26 | F27 | F28 | F29 | F30 |
|---|---|---|---|---|---|---|---|---|---|---|---|---|---|---|---|---|---|---|---|---|---|---|---|---|---|---|---|---|---|---|
| F1 | 1.000 | 0.902 | 0.641 | 0.863 | 0.790 | 0.779 | 0.725 | 0.946 | 0.931 | 0.875 | 0.810 | 0.948 | 0.341 | 0.789 | 0.712 | 0.550 | 0.389 | 0.703 | 0.530 | 0.176 | 0.828 | 0.767 | 0.736 | 0.863 | 0.873 | 0.885 | 0.882 | 0.415 | 0.807 | 0.823 |
| F2 | | 1.000 | 0.468 | 0.937 | 0.620 | 0.832 | 0.890 | 0.986 | 0.951 | 0.972 | 0.940 | 0.986 | 0.172 | 0.624 | 0.548 | 0.451 | 0.585 | 0.873 | 0.768 | 0.032 | 0.985 | 0.888 | 0.901 | 0.994 | 0.997 | 0.999 | 0.954 | 0.016 | 0.485 | 0.829 |
| F3 | | | 1.000 | 0.566 | 0.925 | 0.592 | 0.119 | 0.496 | 0.433 | 0.394 | 0.326 | 0.508 | 0.894 | 0.936 | 0.877 | 0.362 | 0.255 | 0.344 | 0.089 | 0.538 | 0.428 | 0.498 | 0.219 | 0.417 | 0.423 | 0.443 | 0.600 | 0.513 | 0.772 | 0.675 |
| F4 | | | | 1.000 | 0.692 | 0.973 | 0.852 | 0.899 | 0.820 | 0.835 | 0.772 | 0.900 | 0.258 | 0.693 | 0.623 | 0.532 | 0.580 | 0.936 | 0.607 | 0.277 | 0.948 | 0.934 | 0.807 | 0.938 | 0.935 | 0.936 | 0.996 | 0.006 | 0.509 | 0.915 |
| F5 | | | | | 1.000 | 0.692 | 0.330 | 0.651 | 0.586 | 0.543 | 0.470 | 0.659 | 0.684 | 0.999 | 0.978 | 0.488 | 0.033 | 0.452 | 0.154 | 0.357 | 0.569 | 0.666 | 0.307 | 0.581 | 0.578 | 0.594 | 0.718 | 0.551 | 0.846 | 0.785 |
| F6 | | | | | | 1.000 | 0.769 | 0.779 | 0.677 | 0.688 | 0.608 | 0.781 | 0.298 | 0.691 | 0.630 | 0.549 | 0.538 | 0.913 | 0.459 | 0.419 | 0.859 | 0.901 | 0.690 | 0.837 | 0.829 | 0.830 | 0.955 | 0.020 | 0.490 | 0.909 |
| F7 | | | | | | | 1.000 | 0.847 | 0.806 | 0.840 | 0.813 | 0.841 | 0.200 | 0.327 | 0.272 | 0.402 | 0.890 | 0.893 | 0.832 | 0.048 | 0.906 | 0.837 | 0.848 | 0.915 | 0.910 | 0.902 | 0.843 | 0.206 | 0.236 | 0.697 |
| F8 | | | | | | | | 1.000 | 0.986 | 0.981 | 0.947 | 1.000 | 0.203 | 0.654 | 0.575 | 0.453 | 0.522 | 0.801 | 0.726 | 0.017 | 0.942 | 0.826 | 0.881 | 0.965 | 0.974 | 0.980 | 0.920 | 0.159 | 0.584 | 0.792 |
| F9 | | | | | | | | | 1.000 | 0.981 | 0.955 | 0.985 | 0.158 | 0.591 | 0.509 | 0.408 | 0.484 | 0.714 | 0.721 | 0.068 | 0.887 | 0.728 | 0.878 | 0.921 | 0.936 | 0.943 | 0.845 | 0.212 | 0.581 | 0.694 |
| F10 | | | | | | | | | | 1.000 | 0.991 | 0.980 | 0.133 | 0.550 | 0.471 | 0.341 | 0.524 | 0.767 | 0.787 | 0.102 | 0.934 | 0.776 | 0.904 | 0.956 | 0.966 | 0.969 | 0.864 | 0.067 | 0.458 | 0.700 |
| F11 | | | | | | | | | | | 1.000 | 0.945 | 0.091 | 0.478 | 0.403 | 0.257 | 0.507 | 0.723 | 0.811 | 0.180 | 0.908 | 0.729 | 0.890 | 0.928 | 0.939 | 0.940 | 0.806 | 0.013 | 0.373 | 0.629 |
| F12 | | | | | | | | | | | | 1.000 | 0.217 | 0.663 | 0.582 | 0.453 | 0.512 | 0.800 | 0.718 | 0.026 | 0.942 | 0.826 | 0.878 | 0.964 | 0.973 | 0.979 | 0.922 | 0.165 | 0.591 | 0.795 |
| F13 | | | | | | | | | | | | | 1.000 | 0.704 | 0.627 | 0.090 | 0.527 | 0.073 | 0.394 | 0.600 | 0.139 | 0.159 | 0.030 | 0.107 | 0.123 | 0.146 | 0.303 | 0.450 | 0.579 | 0.395 |
| F14 | | | | | | | | | | | | | | 1.000 | 0.974 | 0.478 | 0.041 | 0.454 | 0.152 | 0.363 | 0.574 | 0.665 | 0.317 | 0.585 | 0.583 | 0.599 | 0.721 | 0.543 | 0.840 | 0.783 |
| F15 | | | | | | | | | | | | | | | 1.000 | 0.469 | 0.063 | 0.364 | 0.143 | 0.250 | 0.503 | 0.625 | 0.219 | 0.517 | 0.509 | 0.523 | 0.645 | 0.535 | 0.785 | 0.724 |
| F16 | | | | | | | | | | | | | | | | 1.000 | 0.265 | 0.434 | 0.288 | 0.175 | 0.434 | 0.524 | 0.331 | 0.447 | 0.440 | 0.443 | 0.511 | 0.167 | 0.475 | 0.588 |
| F17 | | | | | | | | | | | | | | | | | 1.000 | 0.717 | 0.715 | 0.117 | 0.629 | 0.603 | 0.609 | 0.635 | 0.624 | 0.607 | 0.548 | 0.382 | 0.064 | 0.413 |
| F18 | | | | | | | | | | | | | | | | | | 1.000 | 0.649 | 0.303 | 0.920 | 0.892 | 0.816 | 0.897 | 0.890 | 0.884 | 0.925 | 0.256 | 0.267 | 0.823 |
| F19 | | | | | | | | | | | | | | | | | | | 1.000 | 0.502 | 0.787 | 0.712 | 0.728 | 0.811 | 0.799 | 0.785 | 0.614 | 0.329 | 0.005 | 0.478 |
| F20 | | | | | | | | | | | | | | | | | | | | 1.000 | 0.045 | 0.126 | 0.023 | 0.001 | 0.009 | 0.020 | 0.265 | 0.216 | 0.390 | 0.343 |
| F21 | | | | | | | | | | | | | | | | | | | | | 1.000 | 0.924 | 0.898 | 0.994 | 0.992 | 0.989 | 0.959 | 0.136 | 0.366 | 0.842 |
| F22 | | | | | | | | | | | | | | | | | | | | | | 1.000 | 0.679 | 0.914 | 0.897 | 0.892 | 0.937 | 0.105 | 0.396 | 0.903 |
| F23 | | | | | | | | | | | | | | | | | | | | | | | 1.000 | 0.893 | 0.908 | 0.908 | 0.814 | 0.207 | 0.245 | 0.649 |
| F24 | | | | | | | | | | | | | | | | | | | | | | | | 1.000 | 0.999 | 0.997 | 0.950 | 0.064 | 0.415 | 0.829 |
| F25 | | | | | | | | | | | | | | | | | | | | | | | | | 1.000 | 0.999 | 0.948 | 0.046 | 0.429 | 0.818 |
| F26 | | | | | | | | | | | | | | | | | | | | | | | | | | 1.000 | 0.951 | 0.022 | 0.451 | 0.822 |
| F27 | | | | | | | | | | | | | | | | | | | | | | | | | | | 1.000 | 0.021 | 0.533 | 0.920 |
| F28 | | | | | | | | | | | | | | | | | | | | | | | | | | | | 1.000 | 0.823 | 0.057 |
| F29 | | | | | | | | | | | | | | | | | | | | | | | | | | | | | 1.000 | 0.600 |
| F30 | | | | | | | | | | | | | | | | | | | | | | | | | | | | | | 1.000 |

TABLE S.IV
THE RANK OF CPI FEATURES AND ACCURACY (%) OF TREE-BASED MODELS ON THE VALIDATION DATASET USING TOP-K OF THE MOST IMPORTANT FEATURES

| RANK | DECISION TREE | | RANDOM FOREST | | EXTRA TREES | | LIGHT GBM | | XGBOOST | | ADABOOST | |
|---|---|---|---|---|---|---|---|---|---|---|---|---|
| | FEATURE | ACCURACY | FEATURE | ACCURACY | FEATURE | ACCURACY | FEATURE | ACCURACY | FEATURE | ACCURACY | FEATURE | ACCURACY |
| 1 | F23 | 73.70 | F23 | 74.36 | F23 | 74.76 | F30 | 64.28 | F13 | 70.36 | F30 | 63.61 |
| 2 | F3 | 78.50 | F30 | 81.06 | F30 | 81.5 | F23 | 77.99 | F3 | 72.45 | F23 | 78.82 |
| 3 | F30 | 84.98 | F3 | 88.19 | F3 | 86.15 | F29 | 86 | F22 | 75.36 | F29 | 84.95 |
| 4 | F22 | 89.23 | F13 | 89.72 | F22 | 88.38 | F1 | 88.78 | F17 | 80.42 | F1 | 93.18 |
| 5 | F29 | 90.92 | F17 | 90.93 | F17 | 90.36 | F28 | 93.31 | F23 | 86.87 | F20 | 95.14 |
| 6 | F28 | 90.63 | F22 | 91.65 | F19 | 89.77 | F9 | 93.35 | F30 | 90.74 | F9 | 94.92 |
| 7 | F1 | 91.99 | F15 | 92.28 | F14 | 89.72 | F20 | 94.2 | F5 | 92.89 | F28 | 94.88 |
| 8 | F19 | 91.48 | F29 | 93.26 | F20 | 91.23 | F22 | 94.35 | F12 | 93.52 | F3 | 95.55 |
| 9 | F15 | 91.96 | F19 | 92.98 | F13 | 90.61 | F3 | 95.04 | F26 | 93.94 | F14 | 95.69 |
| 10 | F13 | 92.09 | F28 | 93.14 | F29 | 92.15 | F14 | 95.17 | F15 | 94.4 | F17 | 95.72 |
| 11 | F5 | 91.76 | F5 | 93.24 | F5 | 92.17 | F5 | 95.17 | F29 | 95.1 | F19 | 95.58 |
| 12 | F9 | 91.83 | F20 | 93.25 | F6 | 91.78 | F19 | 95.11 | F1 | 95.1 | F5 | 95.68 |
| 13 | F20 | 91.93 | F1 | 93.76 | F4 | 91.89 | F15 | 95.3 | F20 | 95.08 | F15 | 95.64 |
| 14 | F17 | 91.64 | F16 | 93.74 | F1 | 92.04 | F16 | 95.39 | F14 | 95.33 | F13 | 95.66 |
| 15 | F14 | 91.67 | F14 | 93.55 | F27 | 92.49 | F6 | 95.35 | F19 | 95.37 | F12 | 95.56 |
| 16 | F16 | 91.81 | F9 | 93.79 | F16 | 92.38 | F2 | 95.46 | F2 | 95.43 | F18 | 95.51 |
| 17 | F24 | 91.81 | F8 | 93.65 | F28 | 92.33 | F17 | 95.2 | F8 | 95.33 | F22 | 95.71 |
| 18 | F11 | 91.64 | F12 | 93.69 | F9 | 92.67 | F18 | 95.17 | F16 | 95.29 | F16 | 95.82 |
| 19 | F12 | 91.65 | F6 | 93.75 | F15 | 92.51 | F12 | 95.34 | F28 | 95.41 | F8 | 95.53 |
| 20 | F6 | 91.47 | F7 | 93.71 | F2 | 92.62 | F13 | 95.18 | F6 | 95.38 | F6 | 95.69 |
| 21 | F18 | 91.44 | F2 | 93.65 | F8 | 92.81 | F8 | 95.06 | F9 | 95.25 | F27 | 95.65 |
| 22 | F8 | 91.52 | F26 | 93.64 | F18 | 92.49 | F26 | 95.06 | F10 | 95.47 | F4 | 95.63 |
| 23 | F25 | 91.49 | F10 | 93.66 | F10 | 92.72 | F7 | 95.23 | F21 | 95.3 | F7 | 95.58 |
| 24 | F21 | 91.73 | F18 | 93.59 | F26 | 92.53 | F21 | 94.84 | F25 | 95.34 | F11 | 95.55 |
| 25 | F7 | 91.61 | F24 | 93.6 | F12 | 92.94 | F27 | 95.43 | F27 | 95.42 | F24 | 95.41 |
| 26 | F10 | 91.54 | F25 | 93.6 | F24 | 92.54 | F11 | 95.23 | F18 | 95.17 | F25 | 95.55 |
| 27 | F2 | 91.6 | F27 | 93.47 | F7 | 92.63 | F24 | 95.48 | F7 | 95.12 | F10 | 95.43 |
| 28 | F4 | 91.52 | F21 | 93.61 | F21 | 92.58 | F25 | 95.24 | F24 | 95.38 | F2 | 95.52 |
| 29 | F26 | 91.49 | F11 | 93.51 | F25 | 92.72 | F4 | 95.13 | F11 | 95.23 | F21 | 95.69 |
| 30 | F27 | 91.54 | F4 | 93.7 | F11 | 92.85 | F10 | 95.23 | F4 | 95.26 | F26 | 95.48 |

## V. CLASSIFICATION ACCURACY FOR SIMULATION DATASETS AND COMPARISON WITH OTHER CPA METHODS

The studied classifiers were tested on two simulation based sets (for $L_1 = 0.4$, $L_2 = 0.5$ and $L_1 = 0.3$, $L_2 = 0.9$). The obtained accuracies are generally very high and similar to the results obtained for the validation dataset (Fig. 4 in the main paper).

TABLE S.V
CLASSIFICATION ACCURACY (%) FOR SIMULATION DATASETS

| CLASSIFICATION ALGORITHM | SIMULATION DATASET $L_1 = 0.4$, $L_2 = 0.5$ | | SIMULATION DATASET $L_1 = 0.3$, $L_2 = 0.9$ | |
|---|---|---|---|---|
| | CONFUSION MATRIX | ACCURACY | CONFUSION MATRIX | ACCURACY |
| Decision Trees | $\begin{bmatrix} 3 & 0 \\ 0 & 32 \end{bmatrix}$ | 100 | $\begin{bmatrix} 1 & 1 \\ 2 & 30 \end{bmatrix}$ | 91.17 |
| Gaussian Naïve Bayes | $\begin{bmatrix} 2 & 1 \\ 3 & 29 \end{bmatrix}$ | 88.57 | $\begin{bmatrix} 1 & 1 \\ 5 & 27 \end{bmatrix}$ | 82.35 |
| Linear Discriminant Analysis | $\begin{bmatrix} 1 & 2 \\ 3 & 29 \end{bmatrix}$ | 85.71 | $\begin{bmatrix} 1 & 1 \\ 1 & 31 \end{bmatrix}$ | 94.11 |
| Light GBM | $\begin{bmatrix} 3 & 0 \\ 1 & 31 \end{bmatrix}$ | 97.14 | $\begin{bmatrix} 1 & 1 \\ 0 & 32 \end{bmatrix}$ | 97.05 |
| XGBoost | $\begin{bmatrix} 3 & 0 \\ 1 & 31 \end{bmatrix}$ | 97.14 | $\begin{bmatrix} 1 & 1 \\ 0 & 32 \end{bmatrix}$ | 97.05 |
| Extra tree | $\begin{bmatrix} 2 & 1 \\ 0 & 32 \end{bmatrix}$ | 97.14 | $\begin{bmatrix} 1 & 1 \\ 1 & 31 \end{bmatrix}$ | 94.11 |
| Random Forest | $\begin{bmatrix} 3 & 0 \\ 0 & 32 \end{bmatrix}$ | 100 | $\begin{bmatrix} 1 & 1 \\ 0 & 32 \end{bmatrix}$ | 97.05 |
| AdaBoost | $\begin{bmatrix} 3 & 0 \\ 0 & 32 \end{bmatrix}$ | 100 | $\begin{bmatrix} 1 & 1 \\ 0 & 32 \end{bmatrix}$ | 97.05 |
| Support Vector Machine | $\begin{bmatrix} 3 & 0 \\ 0 & 32 \end{bmatrix}$ | 100 | $\begin{bmatrix} 1 & 1 \\ 0 & 32 \end{bmatrix}$ | 97.05 |
| k-Nearest Neighbour | $\begin{bmatrix} 3 & 0 \\ 1 & 31 \end{bmatrix}$ | 97.14 | $\begin{bmatrix} 1 & 1 \\ 0 & 32 \end{bmatrix}$ | 97.05 |
| Onln-GFMM | $\begin{bmatrix} 3 & 0 \\ 1 & 31 \end{bmatrix}$ | 97.14 | $\begin{bmatrix} 1 & 1 \\ 1 & 31 \end{bmatrix}$ | 94.11 |
| AGGLO-2 | $\begin{bmatrix} 2 & 1 \\ 0 & 32 \end{bmatrix}$ | 97.14 | $\begin{bmatrix} 1 & 1 \\ 0 & 32 \end{bmatrix}$ | 97.05 |

Generated simulation datasets were used for assessment by other well-known CPA methods: R Index, Idle Index, Area Index, Load disturbance Rejection Performance (LDR) Index and finally, Harris Index. The expected assessment results obtained by R

Index, Idle Index, Area Index and LDR Index based on indices values are presented in Table S. VI. The calculated indices are presented in Table S.VII for ($L_1$ = 0.4, $L_2$ = 0.5 ) and Table S.VIII for ($L_1$ = 0.3, $L_2$ = 0.9) simulation sets, where Score Expert is the expected assessment, based on criteria suggested for generating training dataset for classification algorithm and Score SVM is the output of suggested SVM-based CPA classifier. The results of assessment are also colour-coded based on Table S.VI, where OK is highlighted with green and NOK with red colour. For better clarity, the results are also presented in graphical form in Fig. S4.

TABLE S.VI
ASSESSMENT BASED ON CHOSEN CPA METHODS: R INDEX, IDLE INDEX, AREA INDEX AND LRP INDEX

| R Index | | Idle Index | | Area Index | | LRP Index | |
|---|---|---|---|---|---|---|---|
| NOK (oscillatory) | 1.0 | NOK (sluggish) | 1.0 | NOK (sluggish) | 1.0 | NOK | > 1.4 |
| OK | 0.5 | OK / NOK (oscillatory) | -1.0 | OK | 0.5 | OK | 1.0 |
| NOK (sluggish) | 0.0 | | | NOK (oscillatory) | 0.0 | NOK | < 0.6 |

TABLE S.VII
RESULTS OF ASSESSMENT OF SIMULATION BASED SET ($L_1$ = 0.4, $L_2$ = 0.5)

| Number of response | Score Expert | Score SVM | R Index | Idle Index | Area Index | LRP Index | Harris Index |
|---|---|---|---|---|---|---|---|
| 1 | OK | OK | 0.6270 | -0.1939 | 1.0000 | 0.9887 | 0.3871 |
| 2 | NOK | NOK | 0.5342 | -0.6137 | 0.2740 | 1.1142 | 0.3767 |
| 3 | NOK | NOK | 0.3147 | 0.8930 | 1.0000 | 0.6107 | 0.4650 |
| 4 | NOK | NOK | 0.4462 | 0.6386 | 0.6642 | 0.9179 | 0.4429 |
| 5 | NOK | NOK | 0.5301 | 0.1070 | 1.0000 | 0.6856 | 0.3742 |
| 6 | NOK | NOK | 0.4820 | 0.1329 | 1.0000 | 0.8692 | 0.4328 |
| 7 | NOK | NOK | 0.6383 | -0.7357 | 0.4337 | 1.2327 | 0.3649 |
| 8 | NOK | NOK | 0.3403 | 0.7884 | 0.0872 | 0.7650 | 0.4613 |
| 9 | NOK | NOK | 0.2313 | 0.8772 | 1.0000 | 0.5100 | 0.5009 |
| 10 | NOK | NOK | 1.0690 | -0.4078 | 0.1180 | 0.1339 | 0.0329 |
| 11 | NOK | NOK | 0.9147 | -0.6850 | 0.4050 | 1.3557 | 0.3552 |
| 12 | NOK | NOK | 0.4462 | 0.6386 | 0.6642 | 0.9179 | 0.4429 |
| 13 | NOK | NOK | 0.4379 | 0.1609 | 0.6630 | 0.9179 | 0.4471 |
| 14 | NOK | NOK | 0.4355 | 0.1929 | 1.0000 | 0.7915 | 0.4462 |
| 15 | NOK | NOK | 1.0494 | -0.6838 | 0.1124 | 0.6052 | 0.1742 |
| 16 | NOK | NOK | 0.4937 | -0.7260 | 0.3717 | 1.1246 | 0.3749 |
| 17 | NOK | NOK | 0.9086 | -0.8995 | 0.3734 | 1.1001 | 0.3160 |
| 18 | NOK | NOK | 0.1959 | 0.1500 | 0.2548 | 0.4301 | 0.4384 |
| 19 | NOK | NOK | 0.9472 | -0.6311 | 0.6057 | 1.4707 | 0.3411 |
| 20 | NOK | NOK | 0.9246 | -0.6128 | 0.8035 | 1.3816 | 0.3294 |
| 21 | NOK | NOK | 1.0352 | -0.6692 | 0.4962 | 1.3538 | 0.2849 |
| 22 | NOK | NOK | 1.0208 | -0.6713 | 0.5252 | 1.3639 | 0.2860 |
| 23 | NOK | NOK | 0.4554 | 0.8496 | 0.0518 | 0.8383 | 0.4256 |
| 24 | NOK | NOK | 0.4467 | 0.8626 | 0.0153 | 0.8171 | 0.4278 |
| 25 | OK | OK | 0.5878 | -0.1430 | 0.6564 | 0.9554 | 0.3938 |
| 26 | NOK | NOK | 0.5049 | 0.8188 | 1.0000 | 0.8451 | 0.4099 |
| 27 | NOK | NOK | 0.5658 | -0.0262 | 0.5255 | 1.0327 | 0.4055 |
| 28 | NOK | NOK | 0.6176 | -0.1033 | 0.4705 | 1.0622 | 0.4084 |
| 29 | NOK | NOK | 0.4086 | -0.2740 | 0.5649 | 0.9235 | 0.4510 |
| 30 | NOK | NOK | 0.6576 | -0.6575 | 0.5248 | 1.2462 | 0.3826 |
| 31 | NOK | NOK | 0.5083 | 0.1647 | 0.0177 | 0.8810 | 0.4185 |
| 32 | NOK | NOK | 0.6269 | -0.6490 | 0.4588 | 1.2396 | 0.3962 |
| 33 | NOK | NOK | 0.5287 | 0.6664 | 0.3074 | 0.9263 | 0.4096 |
| 34 | NOK | NOK | 0.4567 | 0.3150 | 1.0000 | 0.4865 | 0.3590 |
| 35 | OK | OK | 0.6012 | 0.6574 | 0.6598 | 0.9449 | 0.3804 |

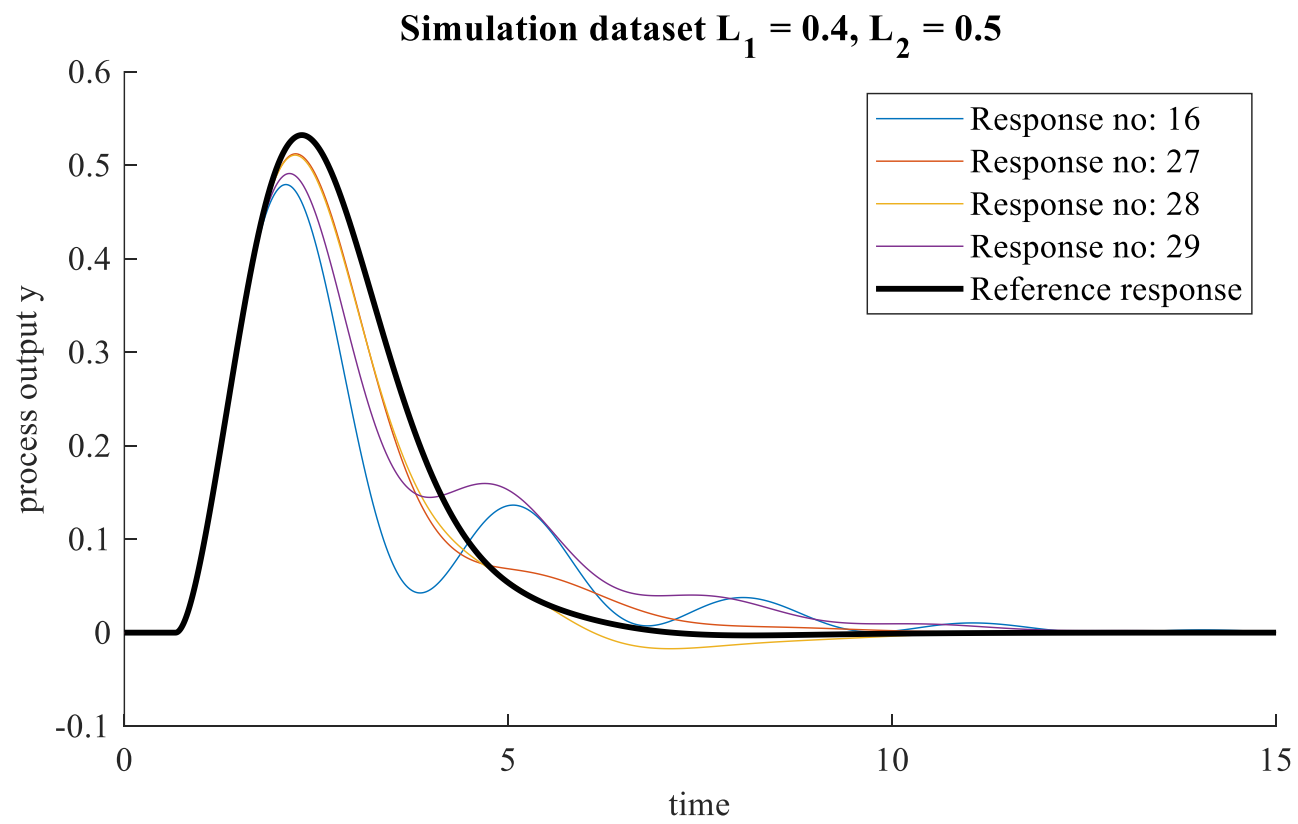

Fig. S3. Process responses, classified as NOK by suggested criteria (Expert) and OK by well-known CPA methods (R Index, Idle Index, Area Index, LRP Index)

TABLE S.VIII

RESULTS OF ASSESSMENT OF SIMULATION BASED SET ($L_1 = 0.3$, $L_2 = 0.9$)

| NUMBER OF RESPONSE | SCORE EXPERT | SCORE SVM | R INDEX | IDLE INDEX | AREA INDEX | LRP INDEX | HARRIS INDEX |
|---|---|---|---|---|---|---|---|
| 1 | OK | OK | 0.5564 | -0.3342 | 1.0000 | 0.9480 | 0.0632 |
| 2 | NOK | NOK | 0.5452 | -0.4309 | 0.6530 | 1.2876 | 0.0762 |
| 3 | NOK | NOK | 0.4469 | 0.8697 | 1.0000 | 0.6439 | 0.0562 |
| 4 | NOK | NOK | 0.5761 | 0.4196 | 0.8982 | 0.9679 | 0.0597 |
| 5 | NOK | NOK | 0.5807 | -0.1245 | 1.0000 | 0.5419 | 0.0431 |
| 6 | NOK | NOK | 0.4487 | 0.8836 | 1.0000 | 0.6629 | 0.0581 |
| 7 | NOK | NOK | 0.6380 | -0.0386 | 0.7204 | 1.1844 | 0.0603 |
| 8 | NOK | NOK | 0.4374 | 0.8596 | 0.4473 | 0.8065 | 0.0650 |
| 9 | NOK | NOK | 0.3185 | 0.8895 | 1.0000 | 0.5377 | 0.0655 |
| 10 | NOK | NOK | 0.8151 | -0.5854 | 0.7924 | 1.5144 | 0.0618 |
| 11 | NOK | NOK | 0.5761 | 0.4196 | 0.8982 | 0.9679 | 0.0597 |
| 12 | OK | NOK | 0.5638 | -0.1041 | 0.9153 | 0.9679 | 0.0610 |
| 13 | NOK | NOK | 0.5907 | -0.2695 | 1.0000 | 0.7664 | 0.0525 |
| 14 | NOK | NOK | 0.5100 | -0.6416 | 0.4977 | 1.3877 | 0.0772 |
| 15 | NOK | NOK | 0.4191 | 0.8592 | 0.2641 | 0.7031 | 0.0600 |
| 16 | NOK | NOK | 0.4369 | 0.8116 | 0.3438 | 0.7031 | 0.0566 |
| 17 | NOK | NOK | 0.2460 | 0.3929 | 1.0000 | 0.4205 | 0.0682 |
| 18 | NOK | NOK | 0.8555 | -0.8055 | 0.8662 | 1.2802 | 0.0517 |
| 19 | NOK | NOK | 0.8346 | -0.8716 | 0.8545 | 1.1781 | 0.0488 |
| 20 | NOK | NOK | 0.9194 | -0.8441 | 0.8137 | 1.3401 | 0.0511 |
| 21 | NOK | NOK | 0.9224 | -0.8046 | 0.8174 | 1.3444 | 0.0514 |
| 22 | NOK | NOK | 0.4275 | 0.8580 | 0.0375 | 0.6380 | 0.0566 |
| 23 | NOK | NOK | 0.4141 | 0.8624 | 1.0000 | 0.6095 | 0.0566 |
| 24 | NOK | NOK | 0.5613 | 0.8012 | 0.7716 | 0.7111 | 0.0493 |
| 25 | NOK | NOK | 0.4979 | 0.8528 | 0.0097 | 0.6292 | 0.0507 |
| 26 | NOK | NOK | 0.5161 | 0.8494 | 0.6189 | 0.7714 | 0.0559 |
| 27 | NOK | NOK | 0.5946 | -0.4869 | 1.0000 | 0.8774 | 0.0556 |
| 28 | NOK | NOK | 0.3504 | 0.8927 | 0.0716 | 0.6902 | 0.0714 |
| 29 | NOK | NOK | 0.5251 | 0.7878 | 0.7306 | 0.9491 | 0.0623 |
| 30 | NOK | NOK | 0.4769 | 0.8548 | 0.0406 | 0.6482 | 0.0532 |
| 31 | NOK | NOK | 0.5032 | 0.7324 | 0.7265 | 0.9491 | 0.0650 |
| 32 | NOK | NOK | 0.5027 | 0.8475 | 0.4358 | 0.7030 | 0.0533 |
| 33 | NOK | NOK | 0.5263 | 0.1553 | 1.0000 | 0.3727 | 0.0382 |
| 34 | NOK | NOK | 0.5712 | 0.7623 | 0.7224 | 0.7171 | 0.0480 |

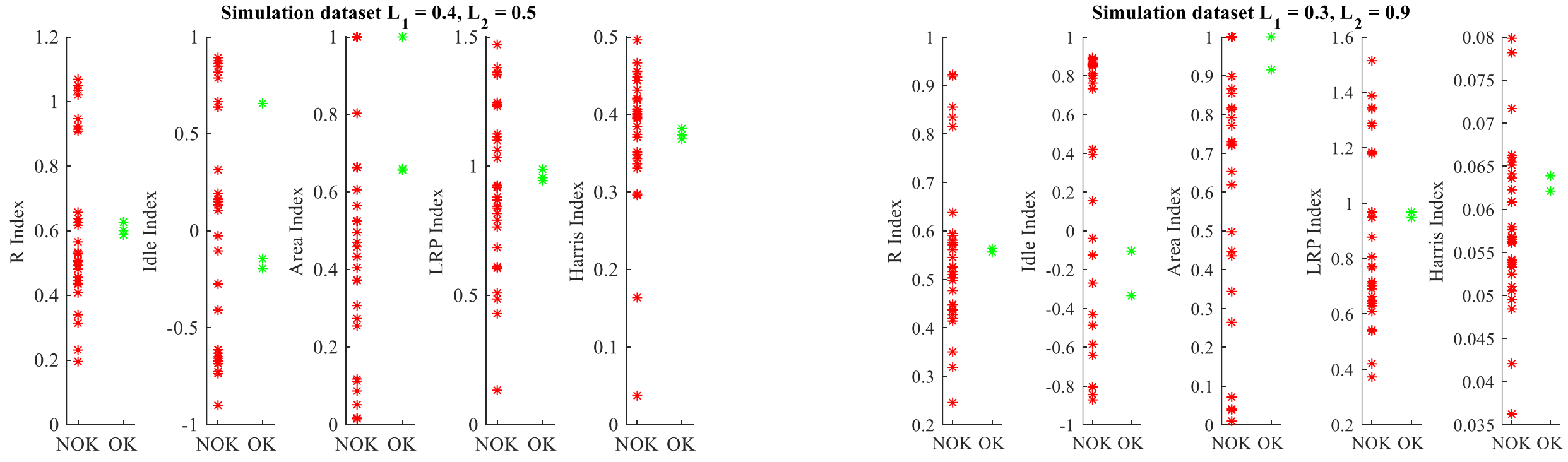

Fig. S4. Graphical representation of calculated performance indices (R Index, Idle Index, Area Index, LRP Index and Harris Index) for OK and NOK samples for simulation sets $L_1 = 0.4$, $L_2 = 0.5$ (left plot) and $L_1 = 0.3$, $L_2 = 0.9$ (right plot).

## VI. SELECTED RESULTS OF SIMULATION BASED VALIDATION

In this section, the results of the simulation based validation of the suggested CPA system based on processes given by the transfer functions (5) and presented in Table I in the main paper are presented. For each process *P1* – *P7*, figures present the accuracy of modelling by FOPDT and SOPDT approximations, comparison of reference disturbance responses for a real process and its SOPDT approximation and a comparison of the reference response with responses classified as OK and NOK.

- Process *P1*.

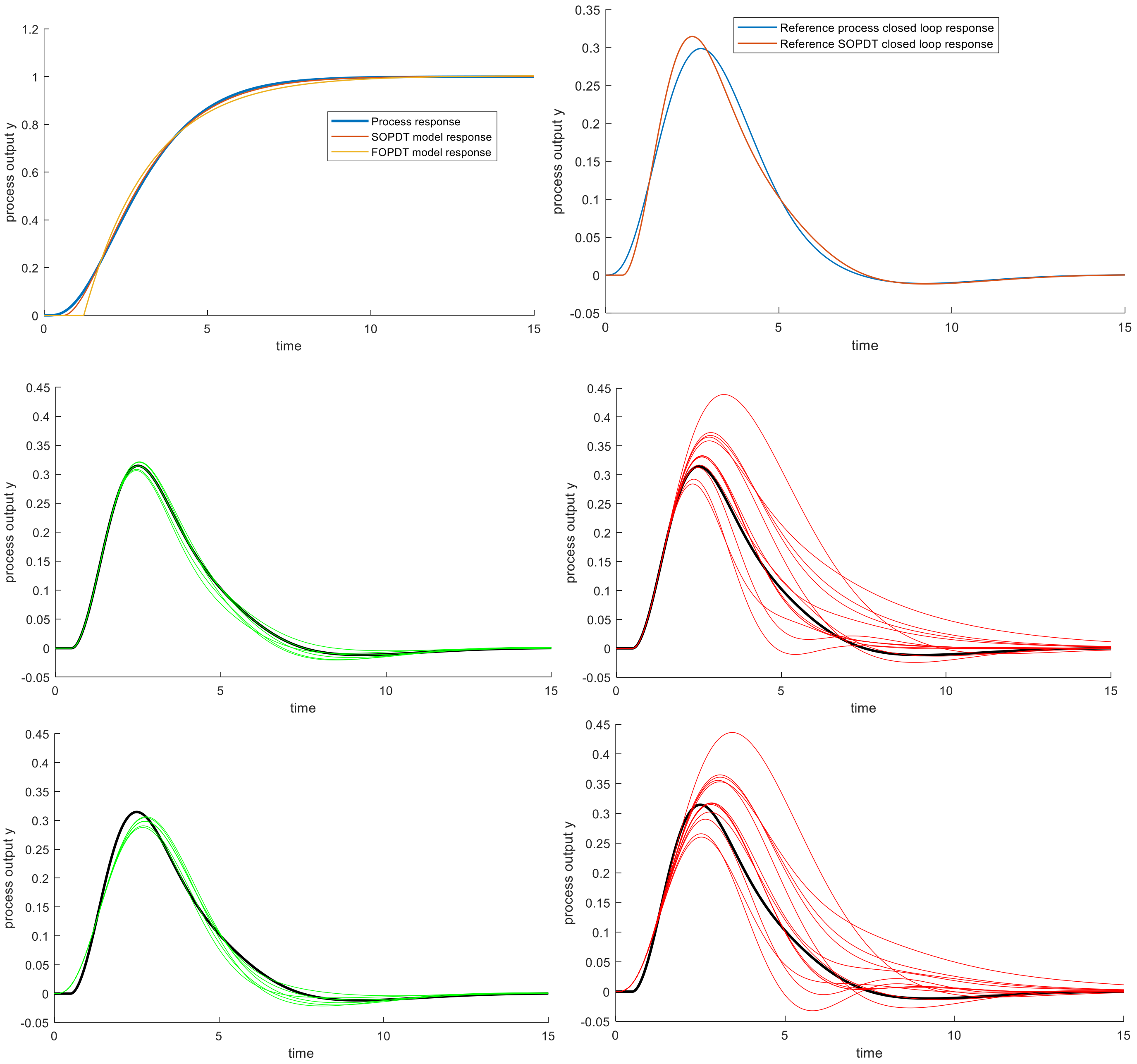

Fig. S5. (Upper row, left). Modelling accuracy for FOPDT and SOPDT approximations. (Upper row, right). Load disturbance rejection responses for the reference PID tunings computed based on the SOPDT approximation. (Middle row). Classification results for the closed loop system with SOPDT process representing an approximation of a real process. (Lower row). Classification results for the closed loop system with a real process. For the middle and lower rows, green colour denotes responses classified as OK and red colour those classified as NOK.

- Process *P2*.

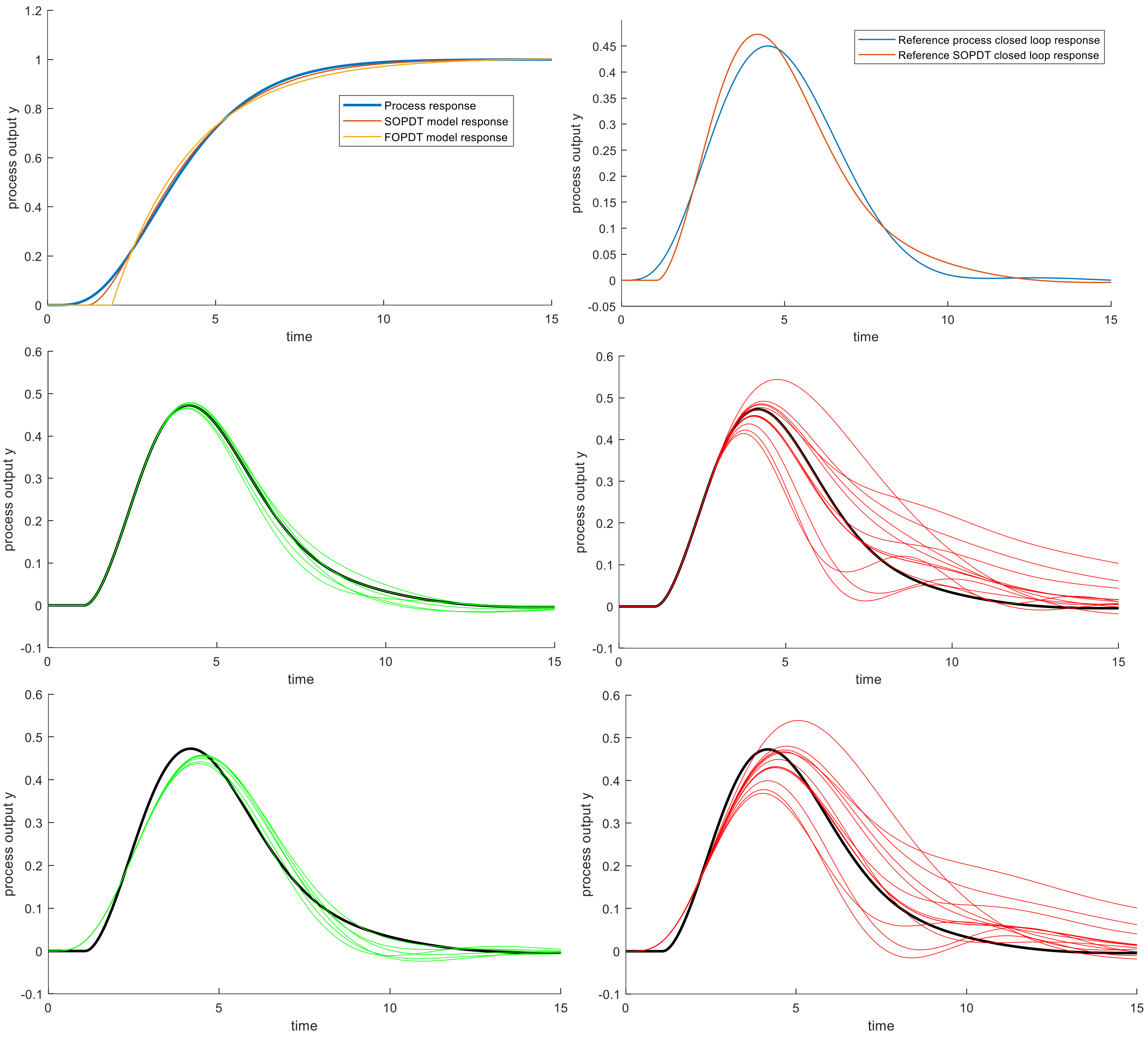

Fig. S6. (Upper row, left). Modelling accuracy for FOPDT and SOPDT approximations. (Upper row, right). Load disturbance rejection responses for the reference PID tunings computed based on the SOPDT approximation. (Middle row). Classification results for the closed loop system with SOPDT process representing an approximation of a real process. (Lower row). Classification results for the closed loop system with a real process. For the middle and lower rows, green colour denotes responses classified as OK and red colour those classified as NOK.

- Process *P3*.

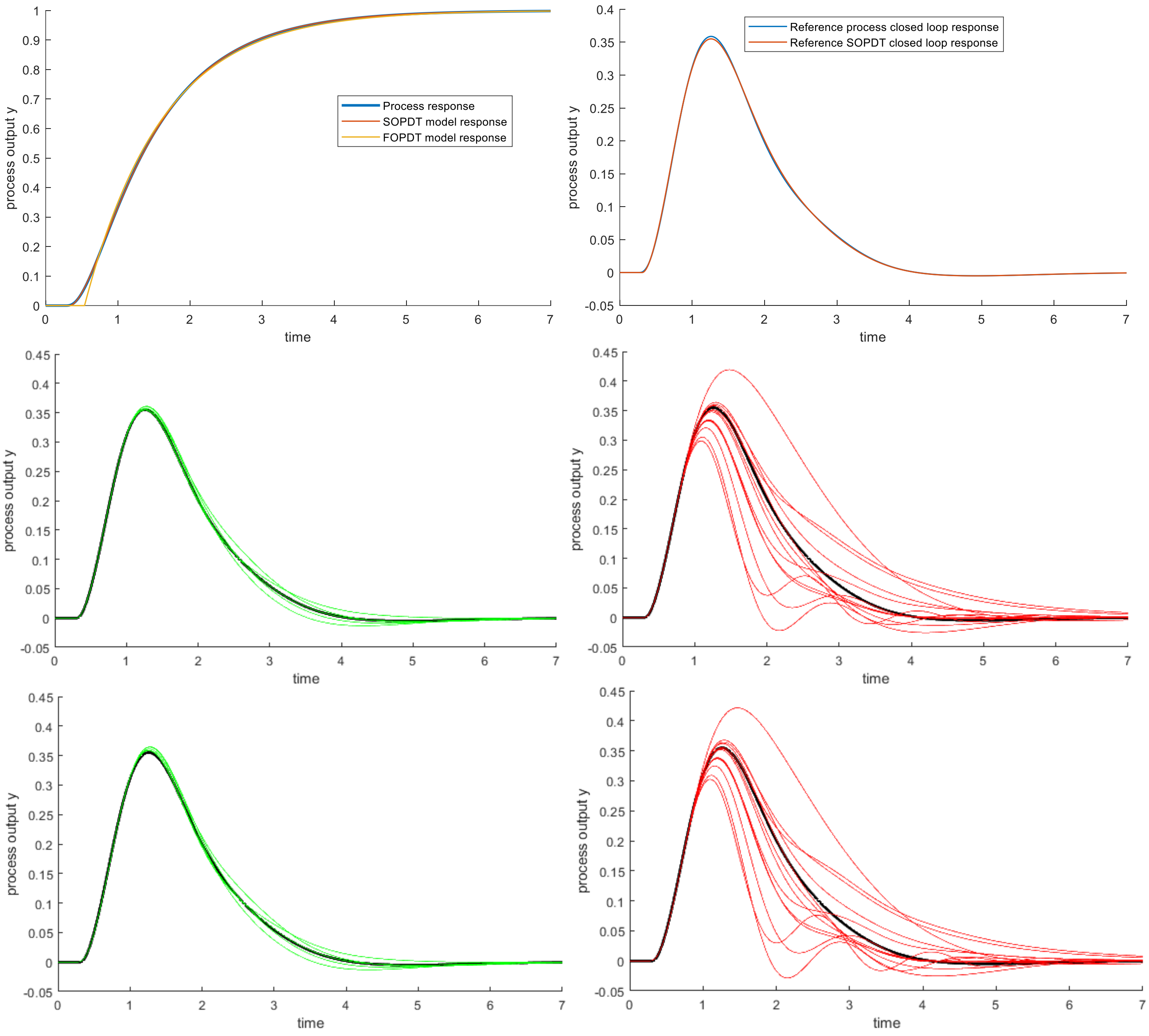

Fig. S7. (Upper row, left). Modelling accuracy for FOPDT and SOPDT approximations. (Upper row, right). Load disturbance rejection responses for the reference PID tunings computed based on the SOPDT approximation. (Middle row). Classification results for the closed loop system with SOPDT process representing an approximation of a real process. (Lower row). Classification results for the closed loop system with a real process. For the middle and lower rows, green colour denotes responses classified as OK and red colour those classified as NOK.

- Process *P4*.

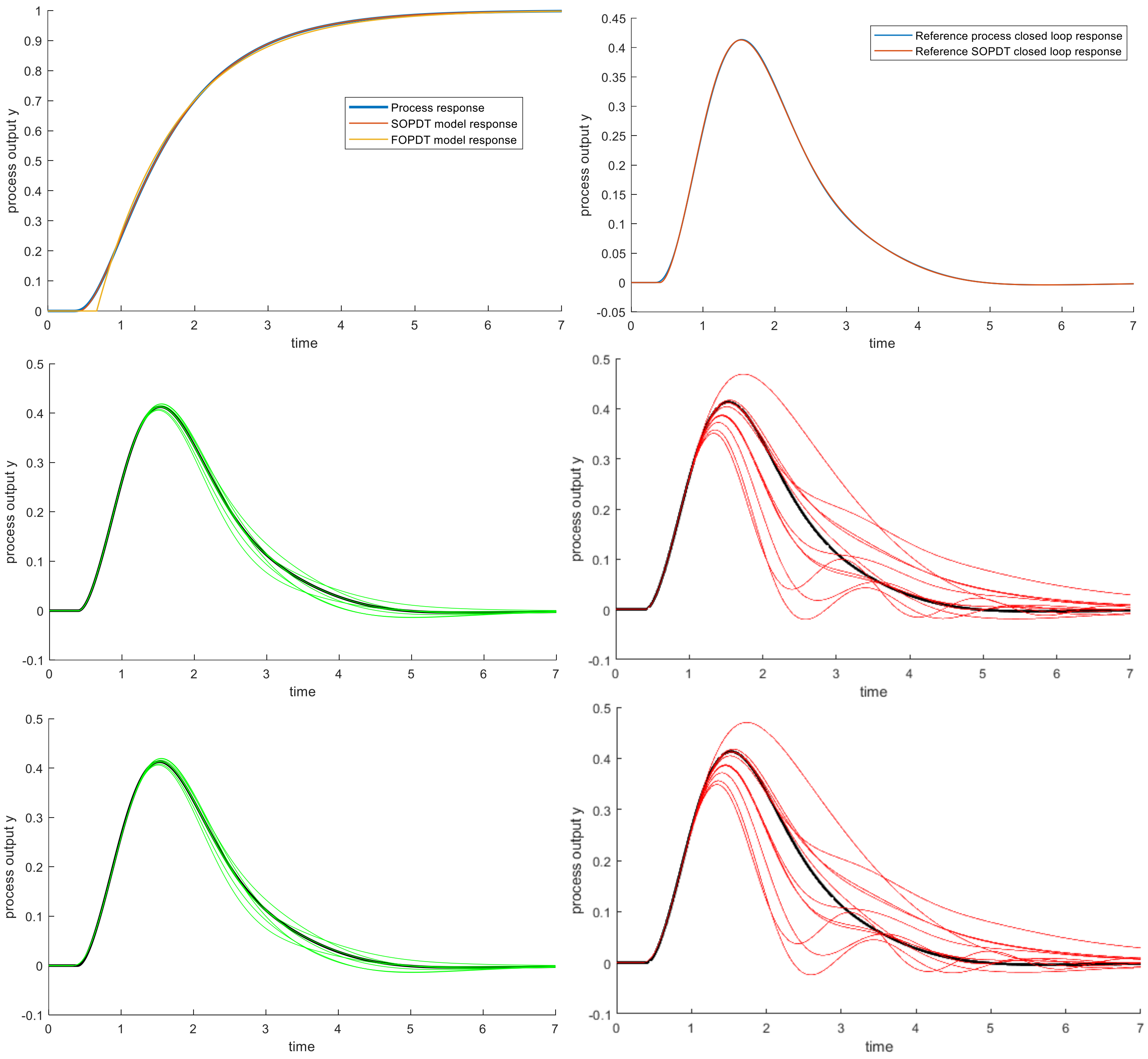

Fig. S8. (Upper row, left). Modelling accuracy for FOPDT and SOPDT approximations. (Upper row, right). Load disturbance rejection responses for the reference PID tunings computed based on the SOPDT approximation. (Middle row). Classification results for the closed loop system with SOPDT process representing an approximation of a real process. (Lower row). Classification results for the closed loop system with a real process. For the middle and lower rows, green colour denotes responses classified as OK and red colour those classified as NOK.

- Process *P5*.

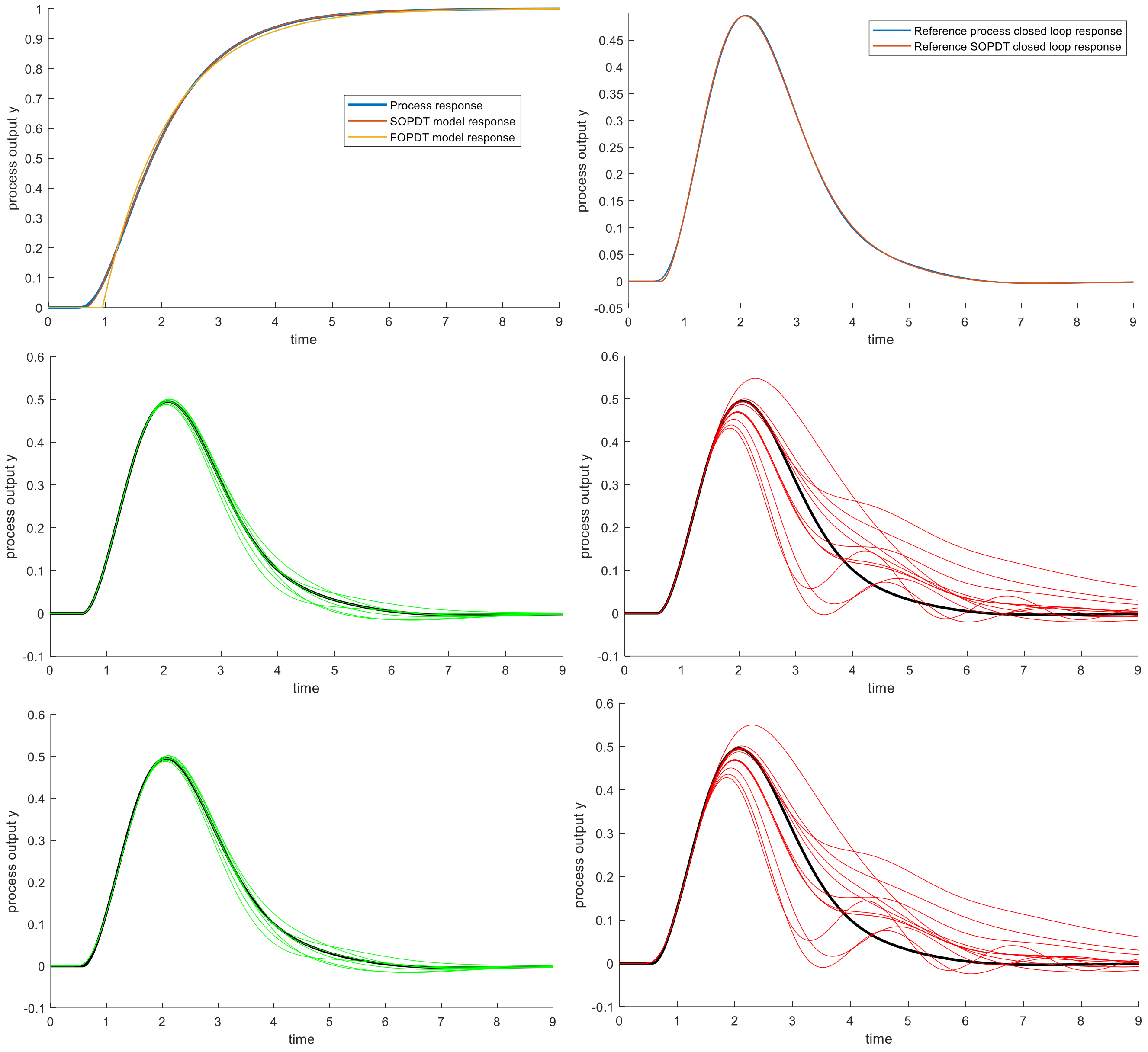

Fig. S9. (Upper row, left). Modelling accuracy for FOPDT and SOPDT approximations. (Upper row, right). Load disturbance rejection responses for the reference PID tunings computed based on the SOPDT approximation. (Middle row). Classification results for the closed loop system with SOPDT process representing an approximation of a real process. (Lower row). Classification results for the closed loop system with a real process. For the middle and lower rows, green colour denotes responses classified as OK and red colour those classified as NOK.

- Process *P6*.

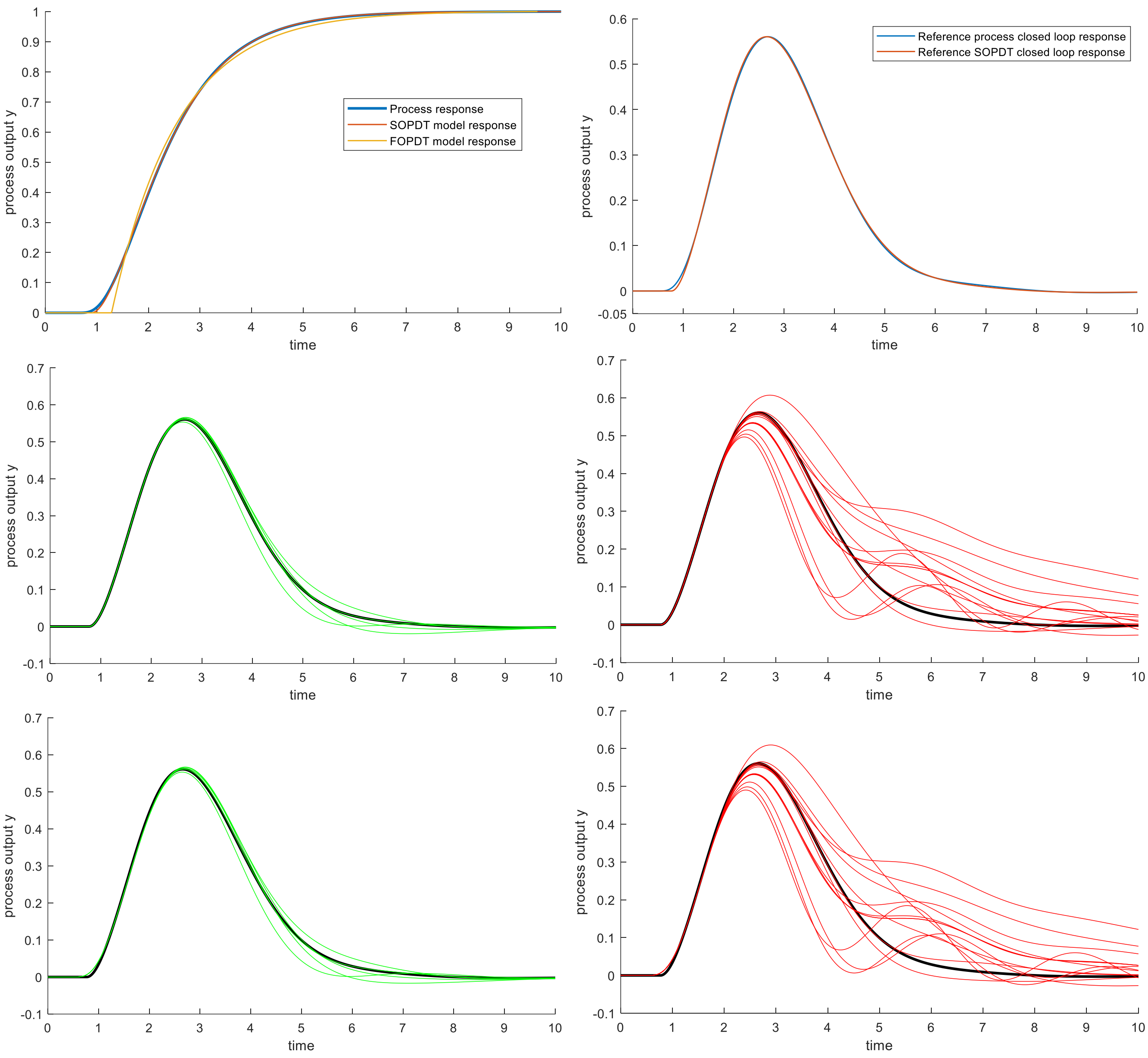

Fig. S10. (Upper row, left). Modelling accuracy for FOPDT and SOPDT approximations. (Upper row, right). Load disturbance rejection responses for the reference PID tunings computed based on the SOPDT approximation. (Middle row). Classification results for the closed loop system with SOPDT process representing an approximation of a real process. (Lower row). Classification results for the closed loop system with a real process. For the middle and lower rows, green colour denotes responses classified as OK and red colour those classified as NOK.

- Process *P7*.

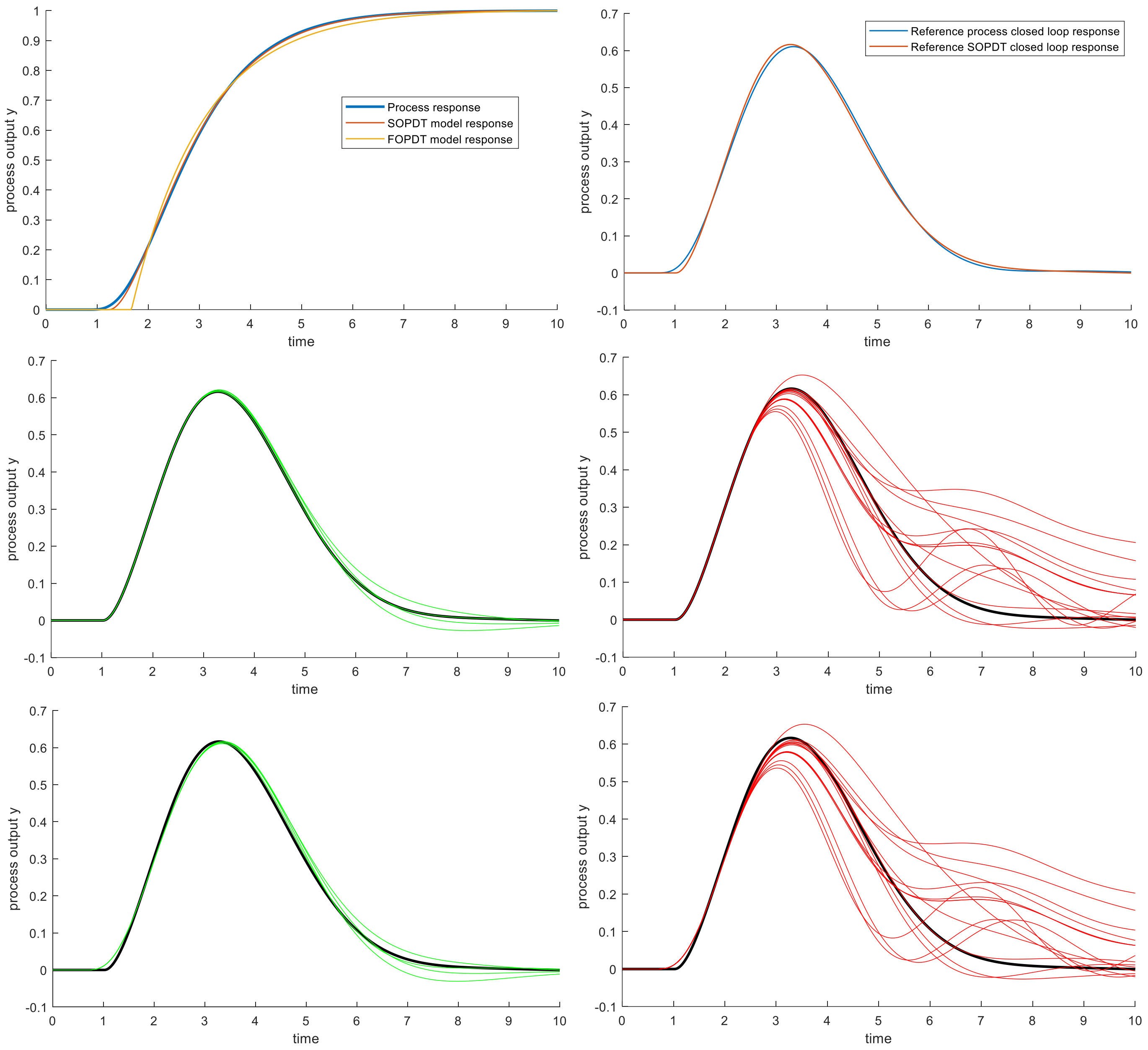

Fig. S11. (Upper row, left). Modelling accuracy for FOPDT and SOPDT approximations. (Upper row, right). Load disturbance rejection responses for the reference PID tunings computed based on the SOPDT approximation. (Middle row). Classification results for the closed loop system with SOPDT process representing an approximation of a real process. (Lower row). Classification results for the closed loop system with a real process. For the middle and lower rows, green colour denotes responses classified as OK and red colour those classified as NOK.

## VII. Details of the cloud-based practical implementation

The example of the practical implementation of the CPA system is intended to assess the current control performance of the PID controller implemented in Siemens S7-1500 Programmable Logic Controller (PLC) during its normal operation. This verification can be performed periodically or upon user request to prevent a significant drop in control performance due to slowly varying fluctuations in process dynamics. In order to prevent PLC from excessive computing load required for CPA functionality, only necessary calculations have been implemented directly in the control program in PLC in the form of dedicated function block "*ControlPerformanceAssessment*". Its application jointly with standard PID Compact function block accessible in TIA Portal is shown in Fig. S12. When CPA procedure is enabled, "*InitializeCPA*" input is set and "*ControlPerformanceAssessment*" function block waits for the steady state that is detected using ICM method [1]. Once the steady state has been detected, a load disturbance step change is applied to the process and its amplitude is adjusted to 10% of the range of manipulating variable stored in the structure connected to the "*PID_CompactConfig*" input. Then, closed loop disturbance rejection response data is collected with sampling time defined by "*SamplingTime*" input until the steady state is detected once again by ICM method after a transient resulting from the process excitation. For monitoring, both steady and transient states are respectively indicated at the outputs "SteadyState" and "*TransientState*". The collected data is stored in PLC's data memory and when this procedure is completed, the data is sent to OPC server jointly with current PID tunings (connected to the input "*PID_CompactCtrlParams*") using a secured OPC UA protocol.

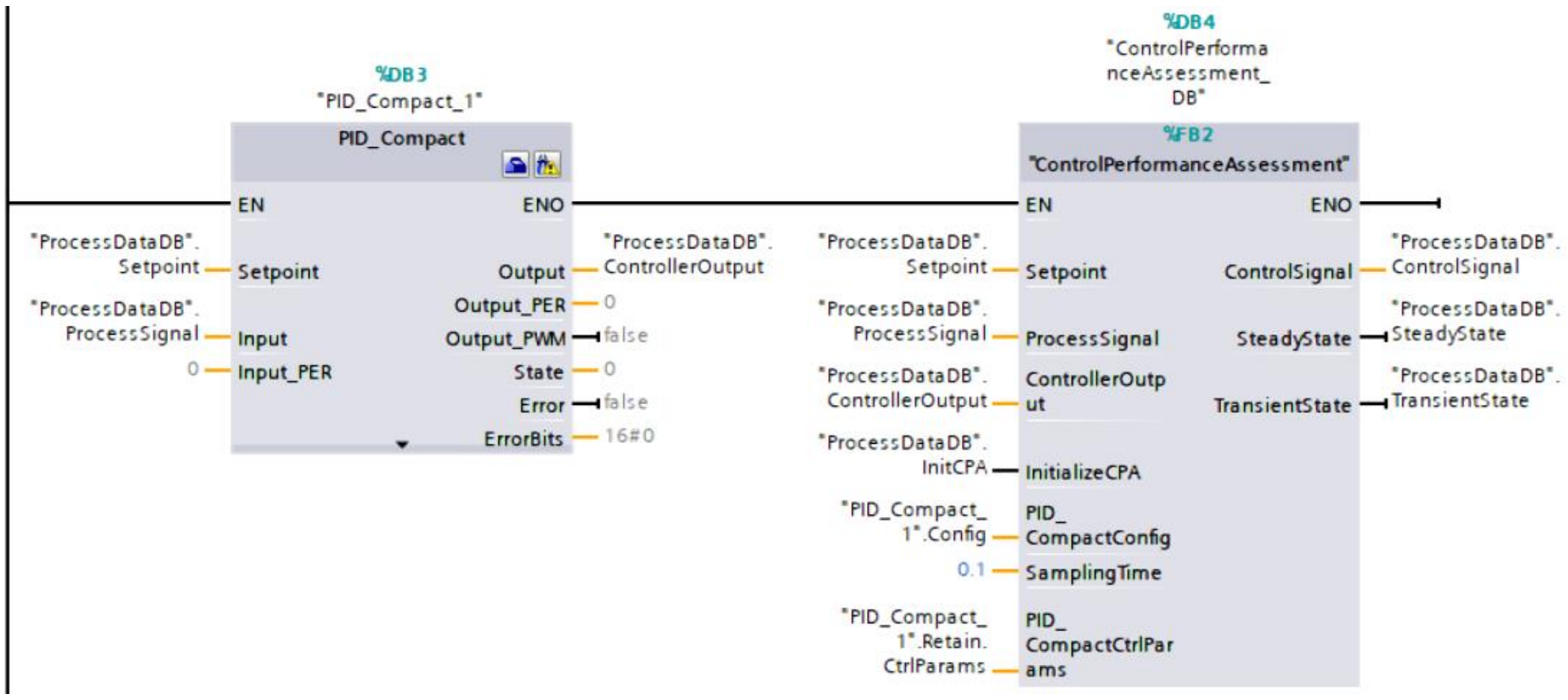

Fig. S12. Siemens S7-1500 PLC-based implementation of "*ControlPerformanceAssessment*" function block in control program.

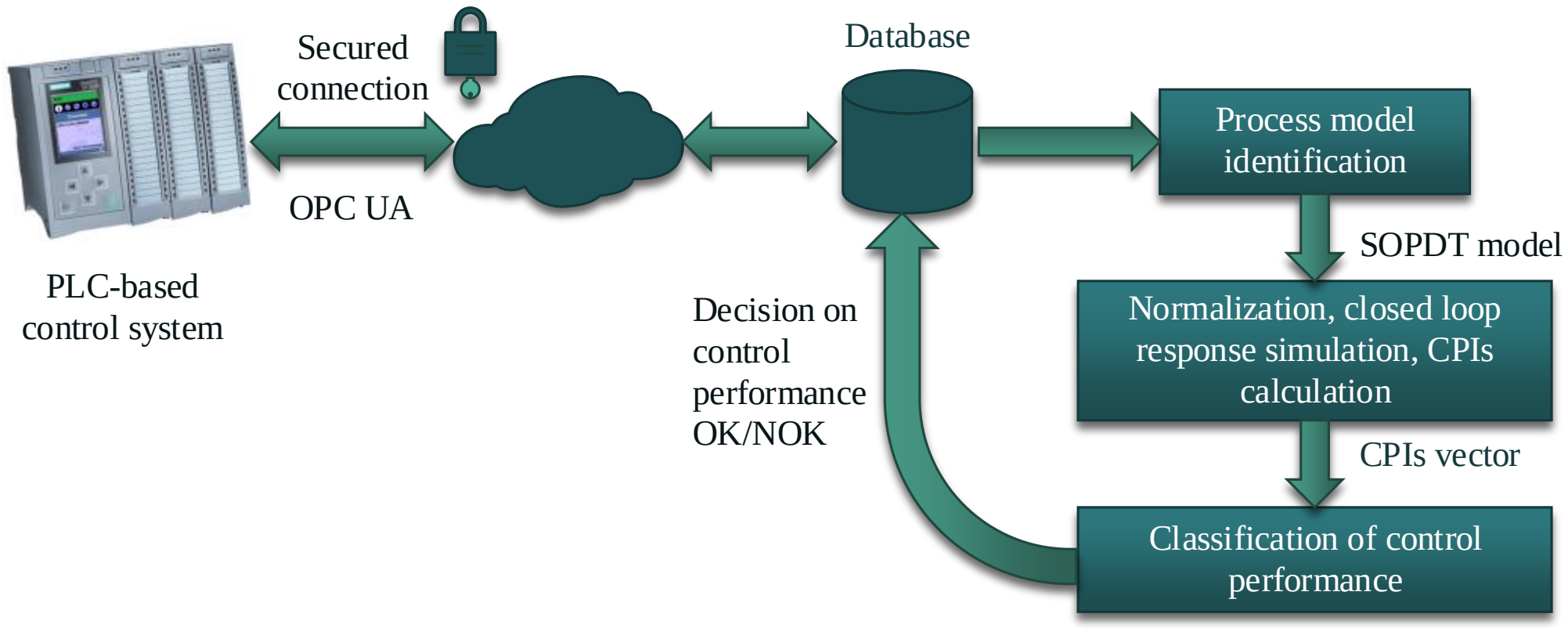

Fig. S13. Architecture of cloud-based implementation of CPA system and its OPC UA connection to PLC-based control system.

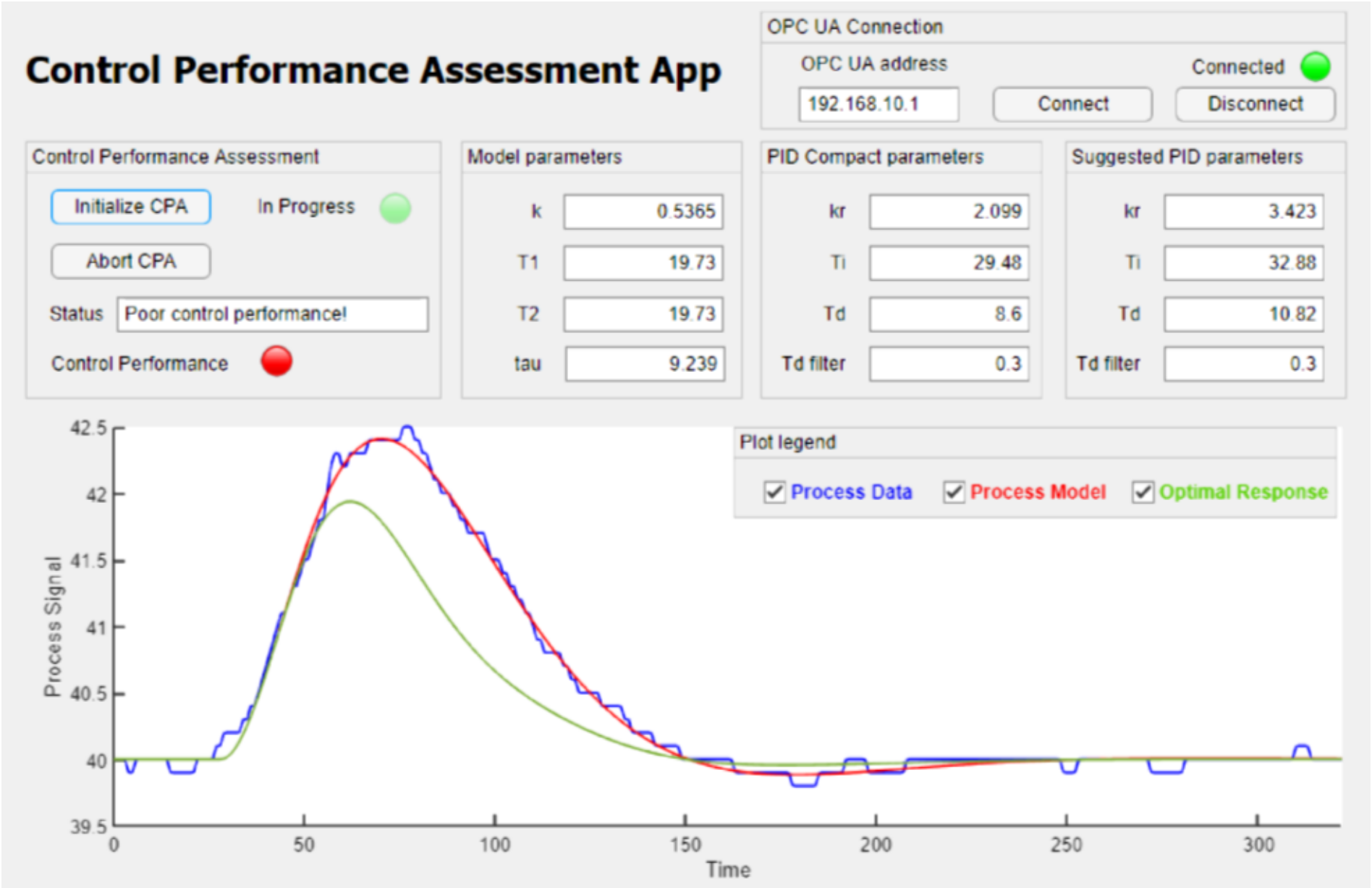

Fig. S14. User interface of exemplary client application for CPA system.

Fig. S13 shows a cloud-based architecture of the considered CPA system. The data collected in PLC is sent to database and based on this data, SOPDT process parameters are identified by nonlinear optimization procedure (Nelder-Mead simplex algorithm) to minimize modelling error. Then, based on identified SOPDT process parameters and the PID tunings, a disturbance rejection response is reconstructed by simulation to minimize the influence of measurement noise. Finally, after computing $L_1$, $L_2$ parameters and appropriate scaling, CPIs are computed for this simulated response as a vector of its features. This is followed by the control performance classification as OK or NOK which is sent to OPC server and then to PLC. It can be also stored in a database and visualized in HMI or SCADA system.

The use of a standard open protocol OPC UA results in full flexibility when it comes to the implementation of client applications. An example of the client user interface application implemented in Matlab is presented in Fig. S14. It provides all essential functionalities, such as connection to OPC UA server, initializing CPA procedure, SOPDT model identification and additionally calculating new PID tunings based on previously identified SOPDT process parameters if the performance was classified as NOK. In cases of uncertain assessment, the user can additionally assess the control performance using the graphical visualization window representing the rejection step response collected from the process by visual comparison with reference rejection response of the assessed control system.

## VIII. List of most important abbreviations and symbols

TABLE S.IX
LIST OF MOST IMPORTANT ABBREVIATIONS

| Abbreviation | Definition |
|---|---|
| CPA | Control Performance Assessment |
| SOPDT | Second Order Plus Delay Time |
| CPIs | Control Performance Indices |
| ML | Machine Learning |
| PID | Proportional Derivative Integral |
| FOPDT | First Order Plus Delay Time |
| IAE | Integral Absolute Error |
| ICM | Increment Count Method |
| LDR Index | Load Disturbance Rejection Performance Index |
| GNB | Gaussian Naïve Bayes |
| LDA | Linear Discriminant Analysis |
| KNN | K-nearest Neighbours |
| DT | Decision Tree |
| GFMM | General Fuzzy Min-Max Neural Network |
| SVM | Support Vector Machine |
| Light GBM | Light Gradient Boosted Machine |
| XGBoost | Extreme Gradient Boosting |
| AdaBoost | Adaptive Boosting |
| Extra Trees | Extremely Randomized Trees |
| RF | Random Forest |
| Onln-GFMM | Online Learning Algorithm for GFMM training |
| AGGLO-2 | Agglomerative Learning Algorithm for GFMM training |
| PLC | Programmable Logic Controller |

TABLE S.X
LIST OF MOST IMPORTANT SYMBOLS

| Symbols | Definition |
|---|---|
| $e$ | Control error |
| $sp$ | Setpoint |
| $y$ | Process variable |
| $\Delta d$ | Step change of load disturbance |
| $k$ | Process gain |
| $\tau_1$, $\tau_2$ | Time constants |
| $\tau_0$ | Delay time |
| $L_1$, $L_2$ | Normalized dynamical parameters |
| $k_r$ | Controller gain (PID parameter) |
| $T_i$ | Integral constant (PID parameter) |
| $T_d$ | Derivative constant (PID parameter) |
| $t_{max}$ | Transient time od closed loop response |
| $A_m$ | Gain margin |
| $\phi_m$ | Phase margin |
| $k_{r,ref}$, $T_{i,ref}$, $T_{d,ref}$ | Reference PID tunings |
| $a_1$, $a_2$, $a_3$ | Randomly generated numbers |
| $N$ | Normal distribution |
| $k_{r,lab}$, $T_{i,lab}$, $T_{d,lab}$ | Modified PID tunings |
| $e_{ref}$ | Reference disturbance rejection response |
| $e_{lab}$ | Considered disturbance rejection response |
| $e_{dist}$ | Normalized distance between disturbance rejection responses of the control system under consideration $e_{lab}$ and reference $e_{ref}$ |
| $A$ | Higher order transfer function parameter |
| $P_h$ | Power of the electric flow heater |
| $T_{in}$ | Inlet temperature |
| $T_{out}$ | Output temperature |
| $T_{SP}$ | Temperature setpoint |
| $F$ | Flow rate |